\documentclass[aps,prb,twocolumn,superscriptaddress,showpacs,floatfix]{revtex4-1}

\usepackage{amsmath}
\usepackage{amssymb}
\usepackage{graphicx}
\usepackage{mathptmx} 
\usepackage{color}  

\begin{document}

\title{An inelastic x-ray study of phonon broadening and charge density wave formation in ortho-II ordered YBa$_{2}$Cu$_{3}$O$_{6.54}$}

\author{E. Blackburn}
\email{e.blackburn@bham.ac.uk}
\affiliation{School of Physics and Astronomy, University of Birmingham, Birmingham B15 2TT, United Kingdom.}

\author{J. Chang}
\affiliation{Institut\,de\,la\,mati\`ere\,complexe,\,Ecole\,Polytechnique\,F\'ederale\,de\,Lausanne (EPFL),\,CH-1015\,Lausanne,\,Switzerland.}

\author{A. H. Said}
\affiliation{Advanced Photon Source, Argonne National Laboratory, Argonne, Illinois 60439, USA}

\author{B. M. Leu}
\affiliation{Advanced Photon Source, Argonne National Laboratory, Argonne, Illinois 60439, USA}

\author{Ruixing Liang}
\affiliation{Department of Physics $\&$ Astronomy, University of British Columbia, Vancouver, Canada}
\affiliation{Canadian Institute for Advanced Research, Toronto, Canada.}

\author{D. A. Bonn}
\affiliation{Department of Physics $\&$ Astronomy, University of British Columbia, Vancouver, Canada}
\affiliation{Canadian Institute for Advanced Research, Toronto, Canada.}

\author{W. N. Hardy}
\affiliation{Department of Physics $\&$ Astronomy, University of British Columbia, Vancouver, Canada}
\affiliation{Canadian Institute for Advanced Research, Toronto, Canada.}

\author{E. M. Forgan}
\affiliation{School of Physics and Astronomy, University of Birmingham, Birmingham B15 2TT, United Kingdom.}

\author{S. M. Hayden}
\email{s.hayden@bris.ac.uk}
\affiliation{H. H. Wills Physics Laboratory, University of Bristol, Bristol, BS8 1TL, United Kingdom.}

\begin{abstract}
Inelastic x-ray scattering is used to investigate charge density wave (CDW) formation and the low-energy lattice dynamics of the underdoped high temperature superconductor ortho-II YBa$_{2}$Cu$_{3}$O$_{6.54}$. We find that, for a temperature $\sim$ 1/3 of the CDW onset temperature ($\approx$ 155 K), the CDW order is static within the resolution of the experiment, that is the inverse lifetime is less than 0.3 meV. In the same temperature region, low-energy phonons near the ordering wavevector of the CDW show large increases in their linewidths. This contrasts with the usual behavior in CDW systems where the phonon anomalies are strongest near the CDW onset temperature.

\end{abstract}


\maketitle

\section{Introduction}

Collective spin or charge density fluctuations are universally present in metals.  In some
materials they remain dynamic. In others, notably the cuprate high temperature superconductors (HTS), they can become static leading to new ordered states.   Charge density waves (CDWs) were recently observed by x-ray diffraction in the high temperature superconductors (HTS) YBa$_2$Cu$_3$O$_{y}$ (YBCO) and (Y/Nd)Ba$_2$Cu$_3$O$_{y}$ \cite{Ghiringhelli2012_GLMB, Chang2012_CBHC, Achkar2012_ASMH, Blackburn2013_BCHH,  Blanco-Canosa2013_BFLL}.  The existence of ground states with competing order is central to many theories of HTS. A widely discussed example is ``stripe order'', that is a state with coexisting charge and spin order \cite{Kivelson2003_KBFO}.

The CDWs observed in YBCO develop inside the pseudogap phase, and compete strongly with HTS, as evidenced by the reduction in CDW Bragg peak intensity observed on cooling through $T_{\mathrm{SC}}$ and the increase in intensity seen on suppressing superconductivity with magnetic field \cite{Chang2012_CBHC, Blackburn2013_BCHH, Blanco-Canosa2013_BFLL}. Signatures of this electronic ordering have been observed by a number of other probes, including scanning probe microscopy  \cite{Hoffman2002_HHLM}, high-field NMR \cite{Wu2011_WMKH,Wu2013_WMKH}, ultrasound \cite{LeBoeuf2013_LKHL}, the Kerr effect \cite{Xia2008_XSDK}, thermal transport \cite{Laliberte2011_LCDH} and the Hall effect \cite{LeBoeuf2011_LDVS}. The transport coefficients are smoothly connected to the low-temperature high-field quantum oscillations (QO) \cite{Doiron-Leyraud2007_DPLL} that demonstrated the existence of small Fermi surface (FS) pockets.  However, some of these measurements \cite{Wu2011_WMKH, LeBoeuf2013_LKHL, Wu2013_WMKH} only show signs of ordering at high fields, in contrast with the x-ray scattering experiments, which observe the CDW in zero field.

The x-ray diffraction measurements reported to date have been performed with no or very broad energy analysis \cite{Ghiringhelli2012_GLMB, Chang2012_CBHC, Achkar2012_ASMH, Blackburn2013_BCHH,  Blanco-Canosa2013_BFLL}. Thus the reported wavevector dependent scans integrate over both Bragg peaks and excitations such as phonons.  As is discussed elsewhere \cite{Ghiringhelli2012_GLMB, Chang2012_CBHC, Blackburn2013_BCHH}, NMR \cite{Wu2011_WMKH} and ultrasound \cite{LeBoeuf2013_LKHL} probe much lower frequencies than conventional x-ray diffraction. It is possible that the correlations involved in the CDW  become ``quasi-static'' (slow) and visible to x-rays before they are seen in NMR and ultrasound.  Thus it is important to make diffraction measurements with higher energy resolution.

In this article, we report a study of the CDW ordering and the low-energy phonons in ortho-II (YBa$_2$Cu$_3$O$_{6.54}$) using inelastic x-ray scattering (IXS) with an energy resolution of 1.5 meV. We find that (i) the CDW is ordered on an energy scale less than the energy resolution of the experiment;  and (ii) there is a strong broadening of the lowest energy phonon linewidth at the ordering wavevector of the CDW, which sets in on entering the CDW state.

\section{Background}
\label{Sec:Background}

Charge density waves are not uncommon in metals (see Ref.~\onlinecite{Monceau2012_Monc} for a recent review). Some of the most well-known examples are the transition metal chalcogenides, such as 2\textit{H}-NbSe$_2$ and TaSe$_2$ \cite{Wilson1975_WiDM}. A charge density wave is essentially the formation of a periodic modulation of the electron density, and is typically associated with a periodic lattice distortion (these may or may not be commensurate with the crystal lattice). This electron (or charge) density modulation may be brought about by electron-phonon or electron-electron interactions.

\begin{figure}[t]
\begin{center}
\includegraphics[width=0.95\linewidth]{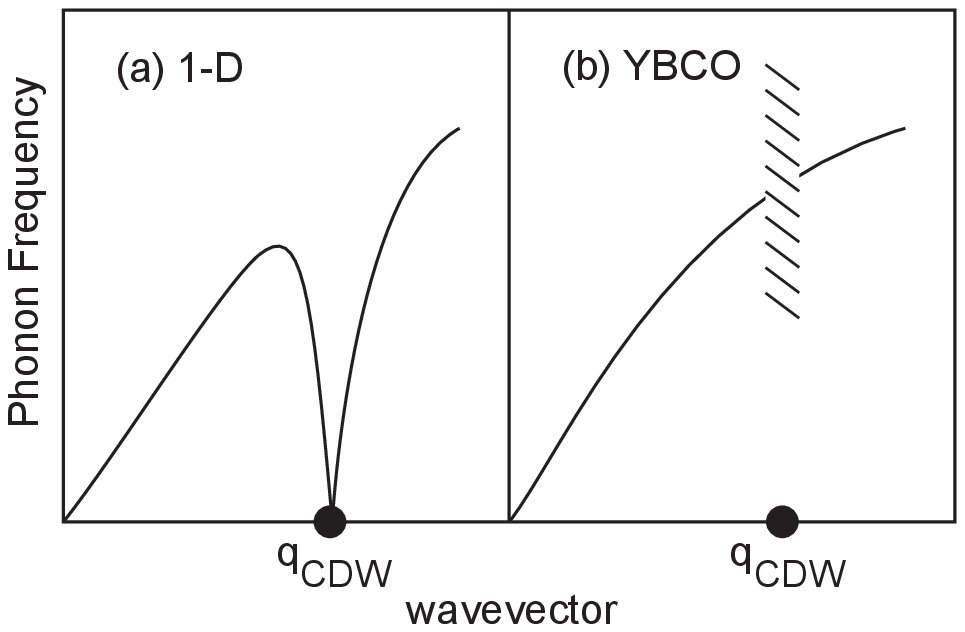}
\end{center}
\caption{ (a) The expected behaviour of the phonon spectrum at $T_{\mathrm{CDW}}$ for a 1D metal, where $q_{\mathrm{CDW}} = 2k_F$; this behaviour is also seen in systems with higher dimensionality e.g.~the 2D CDW in NbSe$_2$ \cite{Wilson1975_WiDM}. (b) The behavior observed in YBa$_2$Cu$_3$O$_{6.54}$ (this work).  The phonon mode does not soften, but there is an increase in damping denoted by the hatched area.}
\label{Fig:phonon_cartoon}
\end{figure}

Structure in the charge response function of a metal is reflected in the phonon energies, via the electron-phonon interaction, this will give rise to Kohn anomalies \cite{Woll1962_WK} in the phonon spectrum associated with the Fermi surface. The response function of a 1-D electron gas diverges at a wavevector 2$k_F$ (twice the Fermi wavevector).  For this case, the Kohn anomaly will occur at a wavevector of 2$k_F$, and if the system exhibits a Peierls instability, the Kohn anomaly will grow, with the phonon frequency eventually softening to zero at a given temperature (Fig.~\ref{Fig:phonon_cartoon}(a)). At this point the CDW has stabilized - the phonon has ``frozen in". This has been observed in 1-D systems such as K$_2$Pt(CN)$_4$Br$_{0.3}$.3.2~D$_2$O (KCP) \cite{Renker1973_RRPG,Carneiro1976_CSWK} and K$_{0.3}$MoO$_3$ \cite{Pouget1991_PHES}.

In higher dimensions, if the same simple physics applies to the electron gas, the renormalization of the phonon frequencies at the Fermi surface is less severe, and a phase transition to the frozen-in state is not expected.  However, strong Kohn anomalies (with $\omega_{\mathrm{ph}} \rightarrow 0$) have also been observed in higher dimensional systems, such as the 2-D system 2$H$-NbSe$_2$ \cite{Wilson1975_WiDM}.  In cases such as these, an additional wavevector dependence must exist, either in the electronic response function $\chi_{\mathbf{q}}$ or in the electron-phonon interaction, and this is sufficient to freeze in the relevant phonon, giving rise to the (static) CDW.

There are also several examples of CDWs developing in systems where $\omega_{\mathrm{ph}} \nrightarrow 0$, e.g. NbSe$_3$ \cite{Requardt2002_RLMC} and (TaSe$_4$)$_2$I \cite{Lorenzo1998_LCMH}. In these materials, there may be some phonon softening, i.e.~a shallow anomaly in the dispersion, and an increase in the phonon linewidth is observed.  In this case, the simple picture of the phonon freezing to give the CDW is not valid.  We note that in all the systems described above, phonon anomalies are strongest near the onset temperature of the CDW, $T_{\mathrm{CDW}}$. In view of the variety of CDW behavior it is important to establish which class of system YBCO belongs to.

\section{Experimental Details}

We carried out (23.724 keV) x-ray scattering experiments at the XOR 30-ID
HERIX beam line \cite{Said2011_SaSD, Toellner2011_ToAS} at the Advanced Photon Source, Argonne National Laboratory.  HERIX allows the x-ray cross-section to be measured as a function of momentum $\mathbf{Q}=\mathbf{k}_i-\mathbf{k}_f$ and energy transfer $E=E_i-E_f$ of the photon.  The scattered beam was analysed by a set of spherically curved silicon (12 12 12) analysers. The full width at half maximum (FWHM) energy resolution was 1.5 meV. The sample was mounted in a closed cycle cryostat on a 4-circle goniometer.  The experiment was performed in transmission geometry.  Inelastic measurements were made in constant wavevector mode. We measured the elastic line during each scan to correct for any drifts in energy calibration.

The sample used was a $\sim$$99\%$ detwinned YBa$_2$Cu$_3$O$_{6.54}$ single crystal of dimensions $1.5 \times 3 \times 0.16$~mm (the same sample used in Ref.~\onlinecite{Blackburn2013_BCHH}).YBa$_2$Cu$_3$O$_{6+x}$ differs from, e.g.~La$_{2-x}$(Ba,Sr)$_{x}$CuO$_4$, in that it contains bilayers of CuO$_2$ planes, separated by layers containing a certain fraction (depending on $x$) of Cu-O chains. The oxygen-filled chains, which run along the orthorhombic crystal {\bf b}-direction tend to order and are labelled ortho-$N$, depending on the repeat length ($Na$) of the ordering of the chains along {\bf a} \cite{Fontaine1987_FoWM, Beyers1989_BAGL, Zimmermann1998_ZVNI}.  This sample was of the type ortho-II, with alternate Cu-O chains occupied.   The lattice parameters are $a \approx 3.82$, $b \approx 3.87$, and $c \approx 11.7$~\AA\ (ignoring the chain-ordering superlattice), $T_{\mathrm{CDW}}=155 \pm 10$~K and $T_{SC}=58$~K (as measured on a Quantum Design MPMS magnetometer).  Following previous practice \cite{Chang2012_CBHC, Blackburn2013_BCHH}, we label reciprocal space using reciprocal lattice units $(2 \pi/a, 2 \pi/b, 2 \pi/c)$ ignoring the additional periodicity introduced by the chain ordering.

\begin{figure}[t]
\begin{center}
\includegraphics[width=0.99\linewidth]{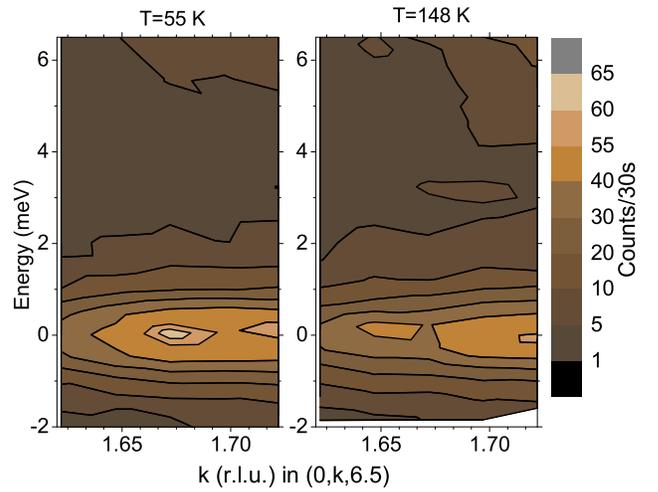}
\end{center}
\caption{Raw intensity plots as a function of $k$ and energy near the $(0,2-\delta_2,6.5)$ position at 55~K $(\approx T_{\mathrm{SC}})$ and 148~K $(\approx T_{\mathrm{CDW}})$, showing the formation of the CDW Bragg peak. The ridge of scattering near zero energy is due to disorder in the sample.  The data have not been corrected for the Bose factor.}
\label{Fig:bragg_contour}
\end{figure}

The CDW produces incommensurate satellite Bragg peaks at positions $\mathbf{Q}=\boldsymbol{\tau} \pm \mathbf{q}_{\mathrm{CDW}}$, where $\boldsymbol{\tau}$ is a reciprocal lattice point of the original (undistorted) lattice and $\mathbf{q}_{\mathrm{CDW}}$ is the wavevector of the CDW.  The CDW in ortho-II YBCO has two fundamental wavevectors, $\mathbf{q}_1=(\delta_1,0,0.5)$ and $\mathbf{q}_2=(0,\delta_2,0.5)$, where $\delta_1=0.320(2)$ and $\delta_2=0.328(2)$ (Ref.~\onlinecite{Blackburn2013_BCHH}).  Previous hard x-ray measurements have indicated that the intensity of the CDW satellite Bragg peaks is strongly dependent on $\boldsymbol{\tau}$, with  $\mathbf{Q}=(0,2-\delta_2,6.5) \approx (0,1.672,6.5)$ being a  particularly strong peak.  We therefore concentrated on measuring near this position.

\section{Results}

\subsection{Charge Ordering}

Fig.~\ref{Fig:bragg_contour} shows contour plots based on a series of constant wavevector scans made in the region of $\mathbf{Q} = (0,2-\delta_2,6.5)$.  The data were collected at $T=55 \pm 2$~K $\approx T_{\mathrm{SC}}$ and $T=148 \pm 2$~K $\approx T_{\mathrm{CDW}}$. A temperature near $T_{\mathrm{SC}}$ was chosen because the previous measurements show that the satellite intensity is maximal at $T_{\mathrm{SC}}$ \cite{Ghiringhelli2012_GLMB, Chang2012_CBHC, Achkar2012_ASMH, Blackburn2013_BCHH,Blanco-Canosa2013_BFLL}.  The $T=148$~K data (right panel) shows a ridge of elastic scattering centered on $E=0$~meV. This background increases towards the (0,~2,~6.5) position and is due to disorder in the sample, frozen in above room temperature. On cooling to $T=55$ K, a peak develops near $\mathbf{Q} \approx (0,1.675,6.5)$ corresponding with the CDW ordering.  This peak is on a sloping background (no background corrections have been made to the data).

\begin{figure}[th]
\begin{center}
\includegraphics[width=0.80\linewidth]{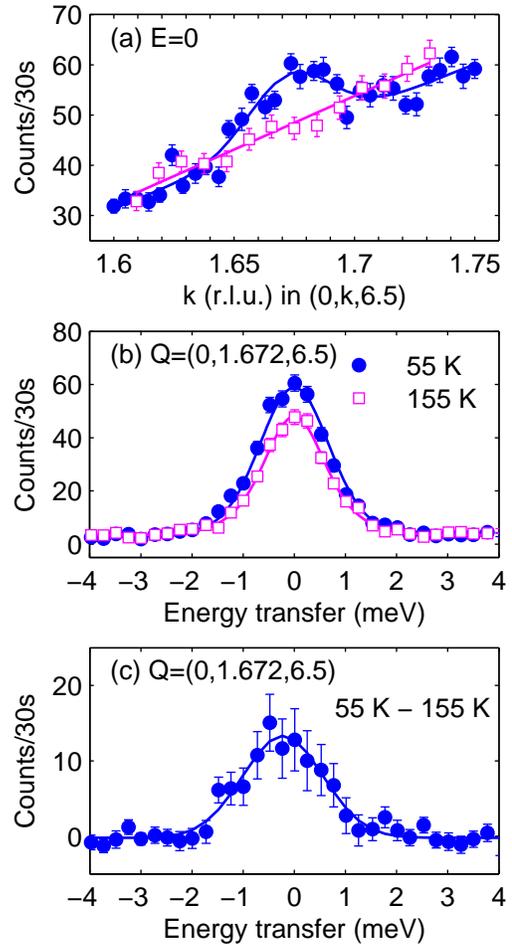}
\end{center}
\caption{(a) $k$-scans and (b) energy scans through the CDW Bragg peak at $T = 55$~K $\approx T_{SC}$ (filled circles) and $T=155$~K $\approx T_{\mathrm{CDW}}$ (open squares). In (a), the wavevector scan at 55~K has been fit as a Gaussian on a sloping background. This gives $\mathbf{q}_2$ = 0.327(2) r.l.u. As is clear from (a), the zero-energy peak at 155 K in (b) is due to background from frozen-in sample disorder. The fits of the energy scans are described in the main text. (c) The difference between the 55~K and 155~K data in (b) fitted to a single Gaussian function.}
\label{Fig:bragg_scans}
\end{figure}

Fig.~\ref{Fig:bragg_scans} shows individual scans through the CDW as a function of wavevector and energy. The value of $\mathbf{q}_2$ measured is 0.327(2) r.l.u., in agreement with the values reported in Ref.~\onlinecite{Blackburn2013_BCHH}. The peak is resolution-limited in the wavevector scans. Comparing the energy dependent scans at 155~K and 55~K, we are able to fit the \textit{additional} scattering due to the CDW Bragg peak using a Gaussian function with a FWHM of $\Delta_{\mathrm{FWHM}}=1.46(10)$~meV. This is indistinguishable from our estimate of the experimental energy resolution (1.5 meV), within experimental error.

\begin{figure}[t]
\begin{center}
\includegraphics[width=0.80\linewidth]{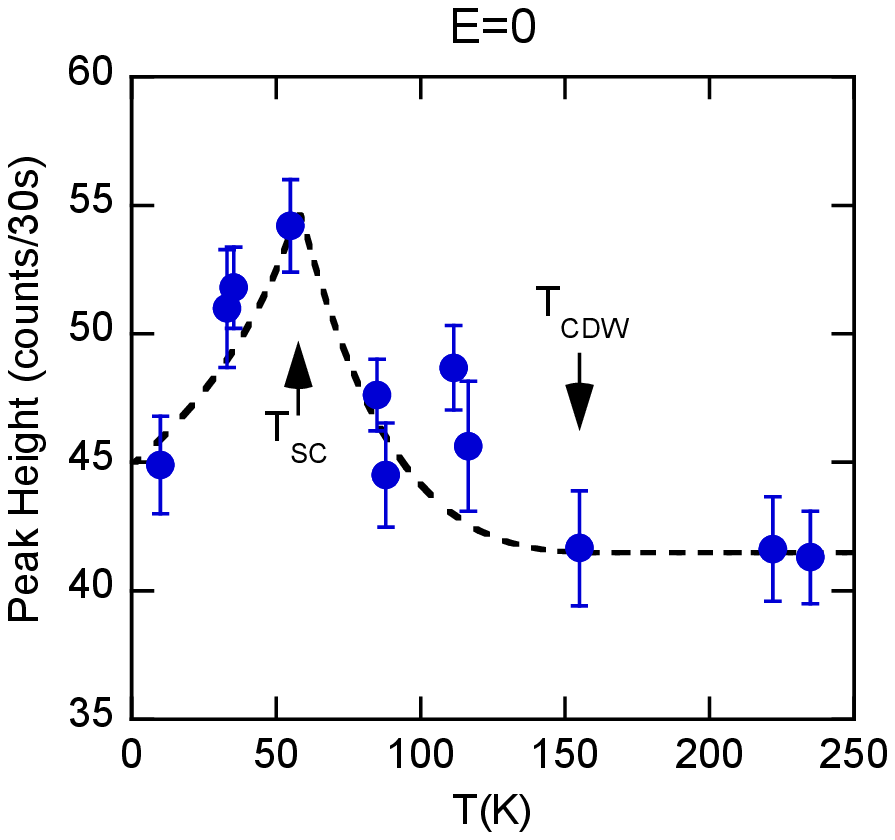}
\end{center}
\caption{Temperature dependence of peak intensity of the CDW satellite Bragg peak at $(0,\delta_2,6.5)$, where $\delta_2=0.327$, determined from energy scans through zero energy, such as in Fig.~\ref{Fig:bragg_scans}(b). The peak height is fitted as described in the text. The dashed line represents the $T$-dependence observed by 100~keV x-ray scattering in Ref.~\onlinecite{Blackburn2013_BCHH}. $T_{\mathrm{SC}}$ and $T_{\mathrm{CDW}}$ are the superconducting and CDW onset temperatures.}
\label{Fig:IvsT}
\end{figure}

We determined the temperature dependence of the CDW Bragg peak from a series of energy dependent scans (as in Fig.~\ref{Fig:bragg_scans}(b)) through the $(0,\delta_2,6.5)$ position. In order to obtain a good fit of the lineshape we used two Gaussian functions centered on $E=0$ (plus a constant background), to provide a better fit of the tails of the peak. For example, the data in Fig.~\ref{Fig:bragg_scans}(b) gave widths $\Delta_1 \approx 1.46$~meV and $\Delta_2 \approx 2.02$~meV.  Fig.~\ref{Fig:IvsT} shows the temperature dependence of the sum of the two peak heights (i.e. the $E=0$ response). The Bragg peak intensity follows our hard x-ray data \cite{Chang2012_CBHC,Blackburn2013_BCHH}.  This was collected without energy analysis, and therefore integrates over a range of energies up to $\sim$ 1 keV, orders of magnitude greater than the $\approx 1.5$~meV resolution of the present measurements. The dotted line in Fig.~\ref{Fig:IvsT} shows the $T$-dependence from our previous experiment, which appears to be consistent with the present experiment. As the temperature is lowered, the Bragg peak intensity starts to increase at $T=155$~K~$\approx T_{\mathrm{CDW}}$. The highest intensity is measured at $T=55$~K~$\approx T_{\mathrm{SC}}$. As the temperature is lowered further the competition between superconductivity and the CDW causes the intensity to reduce \cite{Ghiringhelli2012_GLMB, Chang2012_CBHC, Achkar2012_ASMH,  Blackburn2013_BCHH, Blanco-Canosa2013_BFLL}.

\begin{figure*}[th]
\begin{center}
\includegraphics[width=0.95\linewidth]{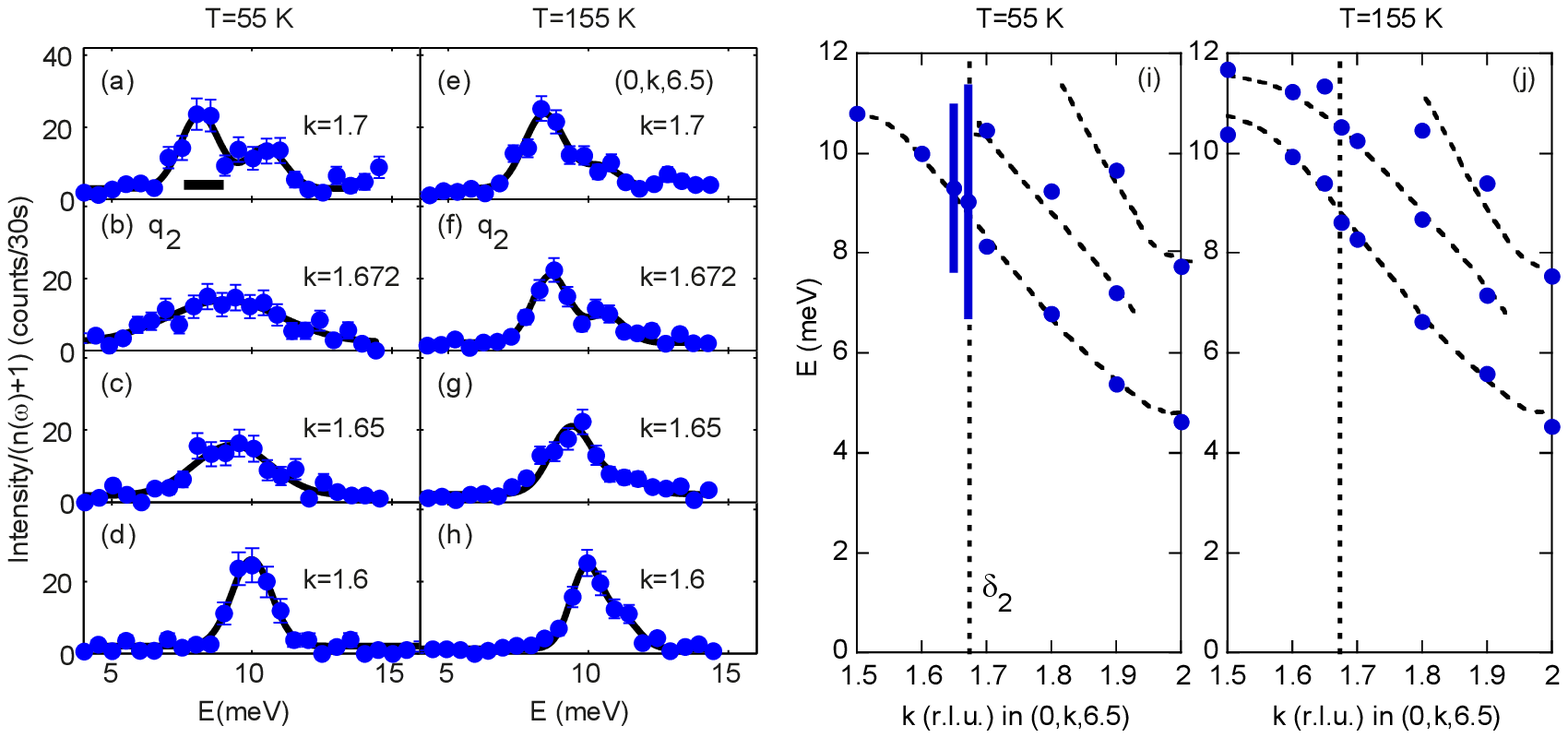}
\end{center}
\caption{(a)-(h) IXS $E$-scans of the low energy phonons for wavevectors along the $(0,k,6.5)$ line. Solid lines are fits to a sum of Gaussian functions. Data have been multiplied by $1-\exp[-E/(k_B T)]$ to correct for the Bose factor. The horizontal bar in panel (a) is the instrumental resolution. (i)-(j) Phonon dispersion curves along the $(0,k,6.5)$ line for $T=55$ and 155~K.  The solid circles represent the phonon peak positions determined from fitting data such as that in (a)-(h); the dashed lines are guides to the eye for the different branches. The resolution-deconvolved phonons widths are represented by vertical bars. The vertical dotted line is the CDW ordering wavevector.}
\label{Fig:phonon_combined}
\end{figure*}

\subsection{Phonon anomalies}

\begin{figure}[t]
\begin{center}
\includegraphics[width=0.80\linewidth]{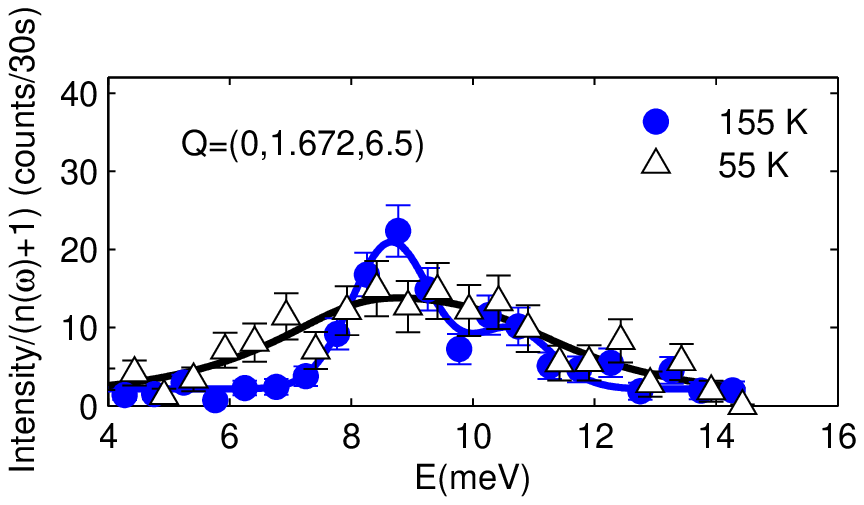}
\end{center}
\caption{IXS $E$-scan of the low-energy phonons at $Q=(0,1.672,6.5)$ for $T=55$ and 155~K. Phonons are fitted to a damped harmonic oscillator (DHO) response function (solid lines). Fits are convolved with the instrumental resolution.}
\label{Fig:phonon_DHO}
\end{figure}

We made energy dependent phonon scans (for $T=55$ and 155~K) at various positions along $(0,k,6.5)$ for $1.5 \le k \le 0$ near the strong $(0,2-\delta_2,6.5)$ CDW satellite Bragg peak found in our previous study \cite{Blackburn2013_BCHH}.
Fig.~\ref{Fig:phonon_combined}(a)-(h) shows energy $E$-scans for wavevectors near the ordering position. Data such as that in Fig.~\ref{Fig:phonon_combined}(a)-(h) may be used to determine the phonon dispersion curves; these are shown in Fig.~\ref{Fig:phonon_combined}(i)-(j).  Fig.~\ref{Fig:phonon_combined} shows that the lowest energy phonons near the ordering wavevector are strongly renomalized on entering the CDW state. Specifically, if we compare the scans at the ordering wavevector $q_2$ - panels (b) and (f) - we see that the line shape of the phonons broadens on cooling from 155~K to 55~K.  This means that spectral weight appears at lower energies (see Fig.~\ref{Fig:phonon_DHO}), and there is a drop in the peak intensity of the phonons.  Similar effects are also observed in panel (c). If we follow the peak intensity across the series (a)-(d), scans (b) and (c) show these anomalies most clearly. The dispersion curves in Fig.~\ref{Fig:phonon_combined}(i)-(j) were obtained by fitting multiple Gaussian lineshapes to the spectra. The points show the peak positions and vertical bars show the resolution-deconvolved (FWHM) widths of the phonon peaks. In most cases, the widths are less than the point size.  However, the big increase in width near $\mathbf{q}_2$ at low temperature is obvious.

Fig.~\ref{Fig:phonon_DHO} shows fits of the data in Fig.~\ref{Fig:phonon_combined}(b,f) to a damped harmonic oscillator response (for each phonon mode),
\begin{eqnarray}
   \chi^{\prime\prime} &\propto&  \frac{\omega \gamma}{(\omega^2-\omega_0^2)^2+(\omega \gamma)^2} \nonumber \\
    &\propto& \frac{1}{2 \omega_1} \left[ \frac{\gamma/2}{(\omega-\omega_1)^2+(\gamma/2)^2}
   -\frac{\gamma/2}{(\omega+\omega_1)^2+(\gamma/2)^2} \right] \label{Eqn:DHO},
 \end{eqnarray}
where $\omega_1^2 = \omega_0^2 - (\gamma/2)^2$. For $T=155~K$, we find the intrinsic phonon widths are zero within the experimental resolution (i.e. $\gamma=0$) and the phonon frequencies, $\hbar \omega_{0,i}$, are $\hbar \omega_{0,1} = 8.7 \pm 0.1$ and $\hbar \omega_{0,2} =10.6 \pm 0.3$~meV. Setting the damping factor $\gamma$ for the two modes to be equal, the response at $T=55$~K can be explained (see Fig.~\ref{Fig:phonon_DHO}) with  $\gamma=4.2 \pm 0.8$~meV and unchanged values for $\hbar \omega_{0,1}$ and $\hbar \omega_{0,2}$. Within this phenomenology, the damping introduces a small shift in the peak of the response (Eqn.~\ref{Eqn:DHO}), $\Delta \omega \approx -\gamma^2/(8 \omega) \approx 0.3$~meV which is not directly discernible within the resolution of with experiment.

\section{Discussion}

\subsection{Charge Ordering}

Charge density waves have rather unique dynamical properties.  It is well known \cite{Monceau2012_Monc} that CDWs can be easily  unpinned from the crystal lattice by the application of an electric field leading to so-called sliding charge density waves (SCDW).\cite{Monceau2012_Monc}  In the case of ortho-II YBCO, the ``charge ordering'' anomalies seen in NMR \cite{Wu2011_WMKH, Wu2013_WMKH} and ultrasound  \cite{LeBoeuf2013_LKHL} occur at lower temperatures and high magnetic fields (Table~\ref{table:charge_order}) than observed with 100~keV x-ray diffraction.

This has led to speculation \cite{LeBoeuf2013_LKHL} that the conventional x-ray diffraction experiments \cite{Ghiringhelli2012_GLMB, Chang2012_CBHC, Achkar2012_ASMH, Blackburn2013_BCHH, Blanco-Canosa2013_BFLL} are, in fact, observing ``dynamic correlations''.  Our IXS measurements are carried out with energy discrimination and therefore allow us to put a bound on the lifetime of the ordered state. We estimate \cite{life_time_note} that, at $T=55$~K, the (deconvolved) FWHM width ($\Delta$) of the CDW Bragg peak is $\Delta < 0.3$~meV.  This puts a lower bound on the lifetime of $\tau \approx 2 \hbar /\Delta \approx  4.4$~ps.  Thus the frequency scale ($\hbar \Gamma_{\mathrm{CDW}} \sim 0.3$~meV) corresponding to the CDW order is at least two orders of magnitude less that than of the superconducting gap $2 \Delta_{\mathrm{SC}} \approx 30$~meV and pseudogap ($ \Delta_{PG} \approx 100$~meV) \cite{Timusk1999_TiSt}.

\begin{table}
\caption{Charge ordering (CO) temperatures in ortho-II YBCO observed by various probes and the energy scale on which the charge correlations are probed ($E_{\mathrm{probe}}$). In the case of NMR and ultrasound $E_{\mathrm{probe}}$ is taken as the energy at which the charge response is probed. For the diffraction measurements, it is the energy range of integration i.e. the energy resolution of the instrument.}
\label{table:charge_order}
\begin{ruledtabular}
\begin{tabular}{llll}
Probe                  & $E_{\mathrm{probe}}$ & $T_{\mathrm{CO}}$(K) & B(T)  \\ \hline
NMR \cite{Wu2011_WMKH} & 1.5~$\mu$eV          & 50                   & 28.5 \\
NMR \cite{Wu2013_WMKH} & 0.5~$\mu$eV          & 60                   & 10.4 \\
ultrasound \cite{LeBoeuf2013_LKHL}& 0.6~$\mu$eV & 44.8               & 26.4 \\
0.931 keV resonant x-ray diffraction\cite{Ghiringhelli2012_GLMB,sample_note} & 130~meV & 150 & 0 \\
100 keV x-ray diffraction \cite{Chang2012_CBHC} & 1~keV & 155(10) & 0 \\
IXS (this work) & 1.5~meV & 150(40) & 0 \\
\end{tabular}
\end{ruledtabular}
\end{table}

Returning to the comparison with NMR \cite{Wu2011_WMKH} and ultrasound \cite{LeBoeuf2013_LKHL} measurement on ortho-II YBCO, these are both low frequency probes, with an energy scale of 1.5 $\mu$eV for the NMR. It is possible that there is an additional phase transition at lower temperatures to a ``frozen'' CDW state seen by these probes. An alternative possibility is a lock-in transition to commensurate ordering, that is, $\delta_2 \rightarrow 1/3$. Such lock-in transitions are common in CDW systems \cite{Wilson1975_WiDM}, and are accompanied by anomalies in the elastic constants \cite{Barmatz1975_BTD} and therefore we deem this more likely.

\subsection{Phonon anomalies}
Charge density waves are closely connected with phonon anomalies. As discussed in Sec.~\ref{Sec:Background}, 1-D CDW systems \cite{Carneiro1976_CSWK,Monceau2012_Monc} show phonon softening or ``Kohn anomalies'', where the frequency of the phonon associated with the CDW is reduced to zero. In less 1-D systems, the anomalies are usually weaker, however, we note the dramatic softening of the phonons recently observed in the well-known CDW material $2H$-NbSe$_{2}$ \cite{Weber2011_WRCO}.

The present results are similar to materials such as NbSe$_3$ \cite{Requardt2002_RLMC} which show a large increase in the phonon linewidth near the CDW ordering wavevector.  The difference between present data on  YBa$_2$Cu$_3$O$_{6.54}$ and other CDW systems \cite{Renker1973_RRPG,Carneiro1976_CSWK,Lorenzo1998_LCMH,Requardt2002_RLMC,Weber2011_WRCO} seems to be the temperature at which the broadening in the phonon linewidths occurs. We observe close to resolution limited linewidths near $T=155$~K $\approx T_{\mathrm{CDW}}$ for the phonons at $\mathbf{q}_2$. These phonons are strongly broadened deep in the CDW state at $T=55$~K $\approx T_{\mathrm{SC}}$.  It is well known \cite{Axe1973_AxSh,Altendorf1993_ACIL,Reznik2012_Rezn} that phonon linewidths (and energies) are affected by superconductivity. On cooling conventional $s$-wave superconductors through $T_{\mathrm{SC}}$, the damping is reduced for low-energy phonons, while for higher energies $\sim 2 \Delta$ it can increase \cite{Axe1973_AxSh,Altendorf1993_ACIL}. This change arises because of the modification of the screening (charge) response of the conduction electrons. The onset of extra damping above $T_{\mathrm{SC}}$ in YBa$_2$Cu$_3$O$_{6.54}$ indicates that it is controlled by something other than the superconducting gap such as the modification of the low-energy electronic properties in the normal state by the pseudogap. Thus, the $d$-wave structure of the pseudogap \cite{Lee2007_LVTL} may mean that the damping is increased for the energy ($\sim 8$~meV) and wavevector [$\mathbf{q}$=(0,0.328,0.5)] of the wavevector of the anomalous phonon studied here.

We note that a new collective low-energy ($\mathbf{Q}$=0) mode, which
appears approximately at $T_{\mathrm{CDW}}$, has been observed by time-resolved
reflectometry \cite{Hinton_HKLV}  in underdoped ortho-III and ortho-VIII YBa$_2$Cu$_3$O$_{6+x}$.
It shows marked changes at $T_{\mathrm{SC}}$.  The mode has frequency $\sim 7.7$~meV,
close to that of the damped phonons reported here. It would also have
components at $\mathbf{q}=\mathbf{q}_{\mathrm{CDW}}$ and therefore may interact with a phonon at this wavevector and energy.

The phonon cross-section of IXS depends on $(\mathbf{Q} \cdot \boldsymbol{\varepsilon})^2$, where $\boldsymbol{\varepsilon}$ is the polarisation of the phonon mode. At the $\mathbf{Q}=(0,2-\delta_2,6.5)$ position, the scattering vector is 52$^{\circ}$ from the $\mathbf{b}^*$ axis, and so our measurement is approximately equally sensitive to atomic motion in the $\mathbf{b}$ and $\mathbf{c}$ directions. Our earlier diffraction work has shown that the CDW modulation has modulations along both these direction, and so we expect that the broadened phonon has displacements along the same directions \cite{Chang2012_CBHC,Blackburn2013_BCHH}.  A recent DFT calculation \cite{Tacon2013_TBSD} suggests the stronger and lower energy mode studied here is the $B_4$, transverse acoustic phonon.

Phonon anomalies have been observed near the $h=0$, $k=0.3$ position at higher energies (40--55~meV) in YBCO \cite{Raichle2011_RRLH, Reznik2008_RPTA, Chung2003_CEMY, Stercel2008_SEMY} and La$_{2-x}$Sr$_x$CuO$_4$ \cite{Pintschovius1999_PiBr, Reznik2012_Rezn}.  These anomalies are associated with oxygen stretching and bending modes, rather than the acoustic mode of the present study. What is interesting is that many phonon branches show an anomaly with the same in-plane component of the wavevector. This suggests that the anomalies stem from the underlying $\mathbf{q}$-dependent electronic dielectric function $\chi(\mathbf{q})$ of the CuO$_2$ planes. Evidence to support this scenario comes from STM measurements on Bi$_2$Sr$_2$CaCu$_2$O$_{8+\delta}$ \cite{Schmidt2011_SFKL, SilvaNeto2012_SAPY} which show modulations in conductance maps with similar wavevectors to the one observed here.  Some studies above \cite{Reznik2008_RPTA,Raichle2011_RRLH} find anomalies to be present in the $k$ direction, but not the $h$ direction -- we do not have the data to comment on this at present.

Chan and Heine \cite{Chan1973_ChHe} have proposed a criterion for a stable CDW phase. Within their picture, CDWs are stabilised either by singularities in the $\mathbf{q}$-dependent electronic dielectric function $\chi(\mathbf{q})$ or by a strongly $\mathbf{q}$-dependent electron-phonon coupling (EPC). Given the evidence for singularities in  $\chi(\mathbf{q})$ presented above, the first scenario of Chan and Heine seems more likely.  This conclusion is strengthened by our observation \cite{Blackburn2013_BCHH} that the $\mathbf{q}$-vectors of the CDWs vary systematically with doping in a manner that indicates a connection with the electronic states near the Fermi surface.

\section{Conclusion}
We have carried out an IXS study of CDW Bragg peak and the low-energy phonon modes in the underdoped cuprate ortho-II YBa$_{2}$Cu$_{3}$O$_{6.54}$.  The CDW order is static within the experimental resolution and we can place an upper bound of 0.3~meV on the energy width of the Bragg peak. This is about two orders of magnitude lower than superconducting gap and the pseudogap.  The lowest energy phonon modes show a strong increase in linewidth near the ordering wavevector of the CDW. This results in the appearance of spectral weight at lower energy transfers on lowering the temperature through the CDW onset temperature.  Unusually, the phonon broadening is largest at $T_{\mathrm{SC}}$ (i.e. $\ll T_{\mathrm{CDW}}$) in contrast with the usual CDW behavior.

\begin{acknowledgments}
We are grateful to Mathieu Le Tacon and collaborators for sharing their extensive IXS data on ortho-VIII YBCO with the community, where strong phonon anomalies are seen at a different reciprocal space position in a sample with different Cu-O chain order. Since the submission of the present paper, this work has appeared as Ref.~\onlinecite{Tacon2013_TBSD}. Our work was supported by the UK EPSRC (grant numbers EP/J015423/1 \& EP/J016977/1), the Swiss National Science Foundation through NCCR-MaNEP and grant number PZ00P2\_142434, the Canadian Natural Sciences and Engineering Research Council and the Canada Foundation for Innovations. Use of the Advanced Photon Source, an Office of Science User Facility operated for the U.S. Department of Energy (DOE) Office of Science by Argonne National Laboratory, was supported by the U.S. DOE under Contract No. DE-AC02-06CH11357. The construction of HERIX was partially supported by the NSF under Grant No. DMR-0115852.
\end{acknowledgments}

\bibliography{ybco_herix}

\begin{thebibliography}{46}%
\makeatletter
\providecommand \@ifxundefined [1]{%
 \@ifx{#1\undefined}
}%
\providecommand \@ifnum [1]{%
 \ifnum #1\expandafter \@firstoftwo
 \else \expandafter \@secondoftwo
 \fi
}%
\providecommand \@ifx [1]{%
 \ifx #1\expandafter \@firstoftwo
 \else \expandafter \@secondoftwo
 \fi
}%
\providecommand \natexlab [1]{#1}%
\providecommand \enquote  [1]{``#1''}%
\providecommand \bibnamefont  [1]{#1}%
\providecommand \bibfnamefont [1]{#1}%
\providecommand \citenamefont [1]{#1}%
\providecommand \href@noop [0]{\@secondoftwo}%
\providecommand \href [0]{\begingroup \@sanitize@url \@href}%
\providecommand \@href[1]{\@@startlink{#1}\@@href}%
\providecommand \@@href[1]{\endgroup#1\@@endlink}%
\providecommand \@sanitize@url [0]{\catcode `\\12\catcode `\$12\catcode
  `\&12\catcode `\#12\catcode `\^12\catcode `\_12\catcode `\%12\relax}%
\providecommand \@@startlink[1]{}%
\providecommand \@@endlink[0]{}%
\providecommand \url  [0]{\begingroup\@sanitize@url \@url }%
\providecommand \@url [1]{\endgroup\@href {#1}{\urlprefix }}%
\providecommand \urlprefix  [0]{URL }%
\providecommand \Eprint [0]{\href }%
\providecommand \doibase [0]{http://dx.doi.org/}%
\providecommand \selectlanguage [0]{\@gobble}%
\providecommand \bibinfo  [0]{\@secondoftwo}%
\providecommand \bibfield  [0]{\@secondoftwo}%
\providecommand \translation [1]{[#1]}%
\providecommand \BibitemOpen [0]{}%
\providecommand \bibitemStop [0]{}%
\providecommand \bibitemNoStop [0]{.\EOS\space}%
\providecommand \EOS [0]{\spacefactor3000\relax}%
\providecommand \BibitemShut  [1]{\csname bibitem#1\endcsname}%
\let\auto@bib@innerbib\@empty
\bibitem [{\citenamefont {Ghiringhelli}\ \emph {et~al.}(2012)\citenamefont
  {Ghiringhelli}, \citenamefont {Le~Tacon}, \citenamefont {Minola},
  \citenamefont {Blanco-Canosa}, \citenamefont {Mazzoli}, \citenamefont
  {Brookes}, \citenamefont {De~Luca}, \citenamefont {Frano}, \citenamefont
  {Hawthorn}, \citenamefont {He}, \citenamefont {Loew}, \citenamefont {Sala},
  \citenamefont {Peets}, \citenamefont {Salluzzo}, \citenamefont {Schierle},
  \citenamefont {Sutarto}, \citenamefont {Sawatzky}, \citenamefont {Weschke},
  \citenamefont {Keimer},\ and\ \citenamefont
  {Braicovich}}]{Ghiringhelli2012_GLMB}%
  \BibitemOpen
  \bibfield  {author} {\bibinfo {author} {\bibfnamefont {G.}~\bibnamefont
  {Ghiringhelli}}, \bibinfo {author} {\bibfnamefont {M.}~\bibnamefont
  {Le~Tacon}}, \bibinfo {author} {\bibfnamefont {M.}~\bibnamefont {Minola}},
  \bibinfo {author} {\bibfnamefont {S.}~\bibnamefont {Blanco-Canosa}}, \bibinfo
  {author} {\bibfnamefont {C.}~\bibnamefont {Mazzoli}}, \bibinfo {author}
  {\bibfnamefont {N.~B.}\ \bibnamefont {Brookes}}, \bibinfo {author}
  {\bibfnamefont {G.~M.}\ \bibnamefont {De~Luca}}, \bibinfo {author}
  {\bibfnamefont {A.}~\bibnamefont {Frano}}, \bibinfo {author} {\bibfnamefont
  {D.~G.}\ \bibnamefont {Hawthorn}}, \bibinfo {author} {\bibfnamefont
  {F.}~\bibnamefont {He}}, \bibinfo {author} {\bibfnamefont {T.}~\bibnamefont
  {Loew}}, \bibinfo {author} {\bibfnamefont {M.~M.}\ \bibnamefont {Sala}},
  \bibinfo {author} {\bibfnamefont {D.~C.}\ \bibnamefont {Peets}}, \bibinfo
  {author} {\bibfnamefont {M.}~\bibnamefont {Salluzzo}}, \bibinfo {author}
  {\bibfnamefont {E.}~\bibnamefont {Schierle}}, \bibinfo {author}
  {\bibfnamefont {R.}~\bibnamefont {Sutarto}}, \bibinfo {author} {\bibfnamefont
  {G.~A.}\ \bibnamefont {Sawatzky}}, \bibinfo {author} {\bibfnamefont
  {E.}~\bibnamefont {Weschke}}, \bibinfo {author} {\bibfnamefont
  {B.}~\bibnamefont {Keimer}}, \ and\ \bibinfo {author} {\bibfnamefont
  {L.}~\bibnamefont {Braicovich}},\ }\href
  {http://www.sciencemag.org/content/337/6096/821.abstract} {\bibfield
  {journal} {\bibinfo  {journal} {Science}\ }\textbf {\bibinfo {volume}
  {337}},\ \bibinfo {pages} {821} (\bibinfo {year} {2012})}\BibitemShut
  {NoStop}%
\bibitem [{\citenamefont {Chang}\ \emph {et~al.}(2012)\citenamefont {Chang},
  \citenamefont {Blackburn}, \citenamefont {Holmes}, \citenamefont
  {Christensen}, \citenamefont {Larsen}, \citenamefont {Mesot}, \citenamefont
  {Liang}, \citenamefont {Bonn}, \citenamefont {Hardy}, \citenamefont
  {Watenphul}, \citenamefont {Zimmermann}, \citenamefont {Forgan},\ and\
  \citenamefont {Hayden}}]{Chang2012_CBHC}%
  \BibitemOpen
  \bibfield  {author} {\bibinfo {author} {\bibfnamefont {J.}~\bibnamefont
  {Chang}}, \bibinfo {author} {\bibfnamefont {E.}~\bibnamefont {Blackburn}},
  \bibinfo {author} {\bibfnamefont {A.~T.}\ \bibnamefont {Holmes}}, \bibinfo
  {author} {\bibfnamefont {N.~B.}\ \bibnamefont {Christensen}}, \bibinfo
  {author} {\bibfnamefont {J.}~\bibnamefont {Larsen}}, \bibinfo {author}
  {\bibfnamefont {J.}~\bibnamefont {Mesot}}, \bibinfo {author} {\bibfnamefont
  {R.}~\bibnamefont {Liang}}, \bibinfo {author} {\bibfnamefont {D.~A.}\
  \bibnamefont {Bonn}}, \bibinfo {author} {\bibfnamefont {W.~N.}\ \bibnamefont
  {Hardy}}, \bibinfo {author} {\bibfnamefont {A.}~\bibnamefont {Watenphul}},
  \bibinfo {author} {\bibfnamefont {M.~v.}\ \bibnamefont {Zimmermann}},
  \bibinfo {author} {\bibfnamefont {E.~M.}\ \bibnamefont {Forgan}}, \ and\
  \bibinfo {author} {\bibfnamefont {S.~M.}\ \bibnamefont {Hayden}},\ }\href
  {http://dx.doi.org/10.1038/nphys2456} {\bibfield  {journal} {\bibinfo
  {journal} {Nat Phys}\ }\textbf {\bibinfo {volume} {8}},\ \bibinfo {pages}
  {871} (\bibinfo {year} {2012})}\BibitemShut {NoStop}%
\bibitem [{\citenamefont {Achkar}\ \emph {et~al.}(2012)\citenamefont {Achkar},
  \citenamefont {Sutarto}, \citenamefont {Mao}, \citenamefont {He},
  \citenamefont {Frano}, \citenamefont {Blanco-Canosa}, \citenamefont
  {Le~Tacon}, \citenamefont {Ghiringhelli}, \citenamefont {Braicovich},
  \citenamefont {Minola}, \citenamefont {Moretti~Sala}, \citenamefont
  {Mazzoli}, \citenamefont {Liang}, \citenamefont {Bonn}, \citenamefont
  {Hardy}, \citenamefont {Keimer}, \citenamefont {Sawatzky},\ and\
  \citenamefont {Hawthorn}}]{Achkar2012_ASMH}%
  \BibitemOpen
  \bibfield  {author} {\bibinfo {author} {\bibfnamefont {A.~J.}\ \bibnamefont
  {Achkar}}, \bibinfo {author} {\bibfnamefont {R.}~\bibnamefont {Sutarto}},
  \bibinfo {author} {\bibfnamefont {X.}~\bibnamefont {Mao}}, \bibinfo {author}
  {\bibfnamefont {F.}~\bibnamefont {He}}, \bibinfo {author} {\bibfnamefont
  {A.}~\bibnamefont {Frano}}, \bibinfo {author} {\bibfnamefont
  {S.}~\bibnamefont {Blanco-Canosa}}, \bibinfo {author} {\bibfnamefont
  {M.}~\bibnamefont {Le~Tacon}}, \bibinfo {author} {\bibfnamefont
  {G.}~\bibnamefont {Ghiringhelli}}, \bibinfo {author} {\bibfnamefont
  {L.}~\bibnamefont {Braicovich}}, \bibinfo {author} {\bibfnamefont
  {M.}~\bibnamefont {Minola}}, \bibinfo {author} {\bibfnamefont
  {M.}~\bibnamefont {Moretti~Sala}}, \bibinfo {author} {\bibfnamefont
  {C.}~\bibnamefont {Mazzoli}}, \bibinfo {author} {\bibfnamefont
  {R.}~\bibnamefont {Liang}}, \bibinfo {author} {\bibfnamefont {D.~A.}\
  \bibnamefont {Bonn}}, \bibinfo {author} {\bibfnamefont {W.~N.}\ \bibnamefont
  {Hardy}}, \bibinfo {author} {\bibfnamefont {B.}~\bibnamefont {Keimer}},
  \bibinfo {author} {\bibfnamefont {G.~A.}\ \bibnamefont {Sawatzky}}, \ and\
  \bibinfo {author} {\bibfnamefont {D.~G.}\ \bibnamefont {Hawthorn}},\ }\href
  {\doibase 10.1103/PhysRevLett.109.167001} {\bibfield  {journal} {\bibinfo
  {journal} {Phys. Rev. Lett.}\ }\textbf {\bibinfo {volume} {109}},\ \bibinfo
  {pages} {167001} (\bibinfo {year} {2012})}\BibitemShut {NoStop}%
\bibitem [{\citenamefont {Blackburn}\ \emph {et~al.}(2013)\citenamefont
  {Blackburn}, \citenamefont {Chang}, \citenamefont {Hucker}, \citenamefont
  {Holmes}, \citenamefont {Christensen}, \citenamefont {Liang}, \citenamefont
  {Bonn}, \citenamefont {Hardy}, \citenamefont {R\"utt}, \citenamefont
  {Gutowski}, \citenamefont {Zimmermann}, \citenamefont {Forgan},\ and\
  \citenamefont {Hayden}}]{Blackburn2013_BCHH}%
  \BibitemOpen
  \bibfield  {author} {\bibinfo {author} {\bibfnamefont {E.}~\bibnamefont
  {Blackburn}}, \bibinfo {author} {\bibfnamefont {J.}~\bibnamefont {Chang}},
  \bibinfo {author} {\bibfnamefont {M.}~\bibnamefont {Hucker}}, \bibinfo
  {author} {\bibfnamefont {A.~T.}\ \bibnamefont {Holmes}}, \bibinfo {author}
  {\bibfnamefont {N.~B.}\ \bibnamefont {Christensen}}, \bibinfo {author}
  {\bibfnamefont {R.}~\bibnamefont {Liang}}, \bibinfo {author} {\bibfnamefont
  {D.~A.}\ \bibnamefont {Bonn}}, \bibinfo {author} {\bibfnamefont {W.~N.}\
  \bibnamefont {Hardy}}, \bibinfo {author} {\bibfnamefont {U.}~\bibnamefont
  {R\"utt}}, \bibinfo {author} {\bibfnamefont {O.}~\bibnamefont {Gutowski}},
  \bibinfo {author} {\bibfnamefont {M.~v.}\ \bibnamefont {Zimmermann}},
  \bibinfo {author} {\bibfnamefont {E.~M.}\ \bibnamefont {Forgan}}, \ and\
  \bibinfo {author} {\bibfnamefont {S.~M.}\ \bibnamefont {Hayden}},\ }\href
  {\doibase 10.1103/PhysRevLett.110.137004} {\bibfield  {journal} {\bibinfo
  {journal} {Phys. Rev. Lett.}\ }\textbf {\bibinfo {volume} {110}},\ \bibinfo
  {pages} {137004} (\bibinfo {year} {2013})}\BibitemShut {NoStop}%
\bibitem [{\citenamefont {Blanco-Canosa}\ \emph {et~al.}(2013)\citenamefont
  {Blanco-Canosa}, \citenamefont {Frano}, \citenamefont {Loew}, \citenamefont
  {Lu}, \citenamefont {Porras}, \citenamefont {Ghiringhelli}, \citenamefont
  {Minola}, \citenamefont {Mazzoli}, \citenamefont {Braicovich}, \citenamefont
  {Schierle}, \citenamefont {Weschke}, \citenamefont {Le~Tacon},\ and\
  \citenamefont {Keimer}}]{Blanco-Canosa2013_BFLL}%
  \BibitemOpen
  \bibfield  {author} {\bibinfo {author} {\bibfnamefont {S.}~\bibnamefont
  {Blanco-Canosa}}, \bibinfo {author} {\bibfnamefont {A.}~\bibnamefont
  {Frano}}, \bibinfo {author} {\bibfnamefont {T.}~\bibnamefont {Loew}},
  \bibinfo {author} {\bibfnamefont {Y.}~\bibnamefont {Lu}}, \bibinfo {author}
  {\bibfnamefont {J.}~\bibnamefont {Porras}}, \bibinfo {author} {\bibfnamefont
  {G.}~\bibnamefont {Ghiringhelli}}, \bibinfo {author} {\bibfnamefont
  {M.}~\bibnamefont {Minola}}, \bibinfo {author} {\bibfnamefont
  {C.}~\bibnamefont {Mazzoli}}, \bibinfo {author} {\bibfnamefont
  {L.}~\bibnamefont {Braicovich}}, \bibinfo {author} {\bibfnamefont
  {E.}~\bibnamefont {Schierle}}, \bibinfo {author} {\bibfnamefont
  {E.}~\bibnamefont {Weschke}}, \bibinfo {author} {\bibfnamefont
  {M.}~\bibnamefont {Le~Tacon}}, \ and\ \bibinfo {author} {\bibfnamefont
  {B.}~\bibnamefont {Keimer}},\ }\href {\doibase
  10.1103/PhysRevLett.110.187001} {\bibfield  {journal} {\bibinfo  {journal}
  {Phys. Rev. Lett.}\ }\textbf {\bibinfo {volume} {110}},\ \bibinfo {pages}
  {187001} (\bibinfo {year} {2013})}\BibitemShut {NoStop}%
\bibitem [{\citenamefont {Kivelson}\ \emph {et~al.}(2003)\citenamefont
  {Kivelson}, \citenamefont {Bindloss}, \citenamefont {Fradkin}, \citenamefont
  {Oganesyan}, \citenamefont {Tranquada}, \citenamefont {Kapitulnik},\ and\
  \citenamefont {Howald}}]{Kivelson2003_KBFO}%
  \BibitemOpen
  \bibfield  {author} {\bibinfo {author} {\bibfnamefont {S.~A.}\ \bibnamefont
  {Kivelson}}, \bibinfo {author} {\bibfnamefont {I.~P.}\ \bibnamefont
  {Bindloss}}, \bibinfo {author} {\bibfnamefont {E.}~\bibnamefont {Fradkin}},
  \bibinfo {author} {\bibfnamefont {V.}~\bibnamefont {Oganesyan}}, \bibinfo
  {author} {\bibfnamefont {J.~M.}\ \bibnamefont {Tranquada}}, \bibinfo {author}
  {\bibfnamefont {A.}~\bibnamefont {Kapitulnik}}, \ and\ \bibinfo {author}
  {\bibfnamefont {C.}~\bibnamefont {Howald}},\ }\href
  {http://link.aps.org/doi/10.1103/RevModPhys.75.1201} {\bibfield  {journal}
  {\bibinfo  {journal} {Rev. Mod. Phys.}\ }\textbf {\bibinfo {volume} {75}},\
  \bibinfo {pages} {1201} (\bibinfo {year} {2003})}\BibitemShut {NoStop}%
\bibitem [{\citenamefont {Hoffman}\ \emph {et~al.}(2002)\citenamefont
  {Hoffman}, \citenamefont {Hudson}, \citenamefont {Lang}, \citenamefont
  {Madhavan}, \citenamefont {Eisaki}, \citenamefont {Uchida},\ and\
  \citenamefont {Davis}}]{Hoffman2002_HHLM}%
  \BibitemOpen
  \bibfield  {author} {\bibinfo {author} {\bibfnamefont {J.~E.}\ \bibnamefont
  {Hoffman}}, \bibinfo {author} {\bibfnamefont {E.~W.}\ \bibnamefont {Hudson}},
  \bibinfo {author} {\bibfnamefont {K.~M.}\ \bibnamefont {Lang}}, \bibinfo
  {author} {\bibfnamefont {V.}~\bibnamefont {Madhavan}}, \bibinfo {author}
  {\bibfnamefont {H.}~\bibnamefont {Eisaki}}, \bibinfo {author} {\bibfnamefont
  {S.}~\bibnamefont {Uchida}}, \ and\ \bibinfo {author} {\bibfnamefont {J.~C.}\
  \bibnamefont {Davis}},\ }\href
  {http://www.sciencemag.org/content/295/5554/466.abstract} {\bibfield
  {journal} {\bibinfo  {journal} {Science}\ }\textbf {\bibinfo {volume}
  {295}},\ \bibinfo {pages} {466} (\bibinfo {year} {2002})}\BibitemShut
  {NoStop}%
\bibitem [{\citenamefont {Wu}\ \emph {et~al.}(2011)\citenamefont {Wu},
  \citenamefont {Mayaffre}, \citenamefont {Kramer}, \citenamefont {Horvatic},
  \citenamefont {Berthier}, \citenamefont {Hardy}, \citenamefont {Liang},
  \citenamefont {Bonn},\ and\ \citenamefont {Julien}}]{Wu2011_WMKH}%
  \BibitemOpen
  \bibfield  {author} {\bibinfo {author} {\bibfnamefont {T.}~\bibnamefont
  {Wu}}, \bibinfo {author} {\bibfnamefont {H.}~\bibnamefont {Mayaffre}},
  \bibinfo {author} {\bibfnamefont {S.}~\bibnamefont {Kramer}}, \bibinfo
  {author} {\bibfnamefont {M.}~\bibnamefont {Horvatic}}, \bibinfo {author}
  {\bibfnamefont {C.}~\bibnamefont {Berthier}}, \bibinfo {author}
  {\bibfnamefont {W.~N.}\ \bibnamefont {Hardy}}, \bibinfo {author}
  {\bibfnamefont {R.}~\bibnamefont {Liang}}, \bibinfo {author} {\bibfnamefont
  {D.~A.}\ \bibnamefont {Bonn}}, \ and\ \bibinfo {author} {\bibfnamefont
  {M.-H.}\ \bibnamefont {Julien}},\ }\href
  {http://dx.doi.org/10.1038/nature10345} {\bibfield  {journal} {\bibinfo
  {journal} {Nature}\ }\textbf {\bibinfo {volume} {477}},\ \bibinfo {pages}
  {191} (\bibinfo {year} {2011})}\BibitemShut {NoStop}%
\bibitem [{\citenamefont {Wu}\ \emph {et~al.}(2013)\citenamefont {Wu},
  \citenamefont {Mayaffre}, \citenamefont {Kramer}, \citenamefont {Horvatic},
  \citenamefont {Berthier}, \citenamefont {Kuhns}, \citenamefont {Reyes},
  \citenamefont {Liang}, \citenamefont {Hardy}, \citenamefont {Bonn},\ and\
  \citenamefont {Julien}}]{Wu2013_WMKH}%
  \BibitemOpen
  \bibfield  {author} {\bibinfo {author} {\bibfnamefont {T.}~\bibnamefont
  {Wu}}, \bibinfo {author} {\bibfnamefont {H.}~\bibnamefont {Mayaffre}},
  \bibinfo {author} {\bibfnamefont {S.}~\bibnamefont {Kramer}}, \bibinfo
  {author} {\bibfnamefont {M.}~\bibnamefont {Horvatic}}, \bibinfo {author}
  {\bibfnamefont {C.}~\bibnamefont {Berthier}}, \bibinfo {author}
  {\bibfnamefont {P.~L.}\ \bibnamefont {Kuhns}}, \bibinfo {author}
  {\bibfnamefont {A.~P.}\ \bibnamefont {Reyes}}, \bibinfo {author}
  {\bibfnamefont {R.}~\bibnamefont {Liang}}, \bibinfo {author} {\bibfnamefont
  {W.~N.}\ \bibnamefont {Hardy}}, \bibinfo {author} {\bibfnamefont {D.~A.}\
  \bibnamefont {Bonn}}, \ and\ \bibinfo {author} {\bibfnamefont {M.-H.}\
  \bibnamefont {Julien}},\ }\href@noop {} {\bibfield  {journal} {\bibinfo
  {journal} {Nature Communications}\ }\textbf {\bibinfo {volume} {4}},\
  \bibinfo {pages} {2113} (\bibinfo {year} {2013})}\BibitemShut {NoStop}%
\bibitem [{\citenamefont {LeBoeuf}\ \emph {et~al.}(2013)\citenamefont
  {LeBoeuf}, \citenamefont {Kramer}, \citenamefont {Hardy}, \citenamefont
  {Liang}, \citenamefont {Bonn},\ and\ \citenamefont
  {Proust}}]{LeBoeuf2013_LKHL}%
  \BibitemOpen
  \bibfield  {author} {\bibinfo {author} {\bibfnamefont {D.}~\bibnamefont
  {LeBoeuf}}, \bibinfo {author} {\bibfnamefont {S.}~\bibnamefont {Kramer}},
  \bibinfo {author} {\bibfnamefont {W.~N.}\ \bibnamefont {Hardy}}, \bibinfo
  {author} {\bibfnamefont {R.}~\bibnamefont {Liang}}, \bibinfo {author}
  {\bibfnamefont {D.~A.}\ \bibnamefont {Bonn}}, \ and\ \bibinfo {author}
  {\bibfnamefont {C.}~\bibnamefont {Proust}},\ }\href
  {http://dx.doi.org/10.1038/nphys2502} {\bibfield  {journal} {\bibinfo
  {journal} {Nat Phys}\ }\textbf {\bibinfo {volume} {9}},\ \bibinfo {pages}
  {79} (\bibinfo {year} {2013})}\BibitemShut {NoStop}%
\bibitem [{\citenamefont {Xia}\ \emph {et~al.}(2008)\citenamefont {Xia},
  \citenamefont {Schemm}, \citenamefont {Deutscher}, \citenamefont {Kivelson},
  \citenamefont {Bonn}, \citenamefont {Hardy}, \citenamefont {Liang},
  \citenamefont {Siemons}, \citenamefont {Koster}, \citenamefont {Fejer},\ and\
  \citenamefont {Kapitulnik}}]{Xia2008_XSDK}%
  \BibitemOpen
  \bibfield  {author} {\bibinfo {author} {\bibfnamefont {J.}~\bibnamefont
  {Xia}}, \bibinfo {author} {\bibfnamefont {E.}~\bibnamefont {Schemm}},
  \bibinfo {author} {\bibfnamefont {G.}~\bibnamefont {Deutscher}}, \bibinfo
  {author} {\bibfnamefont {S.~A.}\ \bibnamefont {Kivelson}}, \bibinfo {author}
  {\bibfnamefont {D.~A.}\ \bibnamefont {Bonn}}, \bibinfo {author}
  {\bibfnamefont {W.~N.}\ \bibnamefont {Hardy}}, \bibinfo {author}
  {\bibfnamefont {R.}~\bibnamefont {Liang}}, \bibinfo {author} {\bibfnamefont
  {W.}~\bibnamefont {Siemons}}, \bibinfo {author} {\bibfnamefont
  {G.}~\bibnamefont {Koster}}, \bibinfo {author} {\bibfnamefont {M.~M.}\
  \bibnamefont {Fejer}}, \ and\ \bibinfo {author} {\bibfnamefont
  {A.}~\bibnamefont {Kapitulnik}},\ }\href@noop {} {\bibfield  {journal}
  {\bibinfo  {journal} {Phys. Rev. Lett.}\ }\textbf {\bibinfo {volume} {100}},\
  \bibinfo {pages} {127002} (\bibinfo {year} {2008})}\BibitemShut {NoStop}%
\bibitem [{\citenamefont {Lalibert\'e}\ \emph {et~al.}(2011)\citenamefont
  {Lalibert\'e}, \citenamefont {Chang}, \citenamefont {Doiron-Leyraud},
  \citenamefont {Hassinger}, \citenamefont {Daou}, \citenamefont {Rondeau},
  \citenamefont {Ramshaw}, \citenamefont {Liang}, \citenamefont {Bonn},
  \citenamefont {Hardy}, \citenamefont {Pyon}, \citenamefont {Takayama},
  \citenamefont {Takagi}, \citenamefont {Sheikin}, \citenamefont {Malone},
  \citenamefont {Proust}, \citenamefont {Behnia},\ and\ \citenamefont
  {L.}}]{Laliberte2011_LCDH}%
  \BibitemOpen
  \bibfield  {author} {\bibinfo {author} {\bibfnamefont {F.}~\bibnamefont
  {Lalibert\'e}}, \bibinfo {author} {\bibfnamefont {J.}~\bibnamefont {Chang}},
  \bibinfo {author} {\bibfnamefont {N.}~\bibnamefont {Doiron-Leyraud}},
  \bibinfo {author} {\bibfnamefont {E.}~\bibnamefont {Hassinger}}, \bibinfo
  {author} {\bibfnamefont {R.}~\bibnamefont {Daou}}, \bibinfo {author}
  {\bibfnamefont {M.}~\bibnamefont {Rondeau}}, \bibinfo {author} {\bibfnamefont
  {B.~J.}\ \bibnamefont {Ramshaw}}, \bibinfo {author} {\bibfnamefont
  {R.}~\bibnamefont {Liang}}, \bibinfo {author} {\bibfnamefont {D.~W.}\
  \bibnamefont {Bonn}}, \bibinfo {author} {\bibfnamefont {W.~N.}\ \bibnamefont
  {Hardy}}, \bibinfo {author} {\bibfnamefont {S.}~\bibnamefont {Pyon}},
  \bibinfo {author} {\bibfnamefont {T.}~\bibnamefont {Takayama}}, \bibinfo
  {author} {\bibfnamefont {H.}~\bibnamefont {Takagi}}, \bibinfo {author}
  {\bibfnamefont {L.}~\bibnamefont {Sheikin}}, \bibinfo {author} {\bibfnamefont
  {I.}~\bibnamefont {Malone}}, \bibinfo {author} {\bibfnamefont
  {C.}~\bibnamefont {Proust}}, \bibinfo {author} {\bibfnamefont
  {K.}~\bibnamefont {Behnia}}, \ and\ \bibinfo {author} {\bibfnamefont
  {T.}~\bibnamefont {L.}},\ }\href@noop {} {\bibfield  {journal} {\bibinfo
  {journal} {Nature Communications}\ }\textbf {\bibinfo {volume} {2}},\
  \bibinfo {pages} {432} (\bibinfo {year} {2011})}\BibitemShut {NoStop}%
\bibitem [{\citenamefont {LeBoeuf}\ \emph {et~al.}(2011)\citenamefont
  {LeBoeuf}, \citenamefont {Doiron-Leyraud}, \citenamefont {Vignolle},
  \citenamefont {Sutherland}, \citenamefont {Ramshaw}, \citenamefont
  {Levallois}, \citenamefont {Daou}, \citenamefont {F.}, \citenamefont
  {Cyr-Choiniere}, \citenamefont {Chang}, \citenamefont {Jo}, \citenamefont
  {Balicas}, \citenamefont {Liang}, \citenamefont {Bonn}, \citenamefont
  {Hardy}, \citenamefont {Proust},\ and\ \citenamefont
  {Taillefer}}]{LeBoeuf2011_LDVS}%
  \BibitemOpen
  \bibfield  {author} {\bibinfo {author} {\bibfnamefont {D.}~\bibnamefont
  {LeBoeuf}}, \bibinfo {author} {\bibfnamefont {N.}~\bibnamefont
  {Doiron-Leyraud}}, \bibinfo {author} {\bibfnamefont {B.}~\bibnamefont
  {Vignolle}}, \bibinfo {author} {\bibfnamefont {M.}~\bibnamefont
  {Sutherland}}, \bibinfo {author} {\bibfnamefont {B.~J.}\ \bibnamefont
  {Ramshaw}}, \bibinfo {author} {\bibfnamefont {J.}~\bibnamefont {Levallois}},
  \bibinfo {author} {\bibfnamefont {R.}~\bibnamefont {Daou}}, \bibinfo {author}
  {\bibfnamefont {L.}~\bibnamefont {F.}}, \bibinfo {author} {\bibfnamefont
  {O.}~\bibnamefont {Cyr-Choiniere}}, \bibinfo {author} {\bibfnamefont
  {J.}~\bibnamefont {Chang}}, \bibinfo {author} {\bibfnamefont {Y.~J.}\
  \bibnamefont {Jo}}, \bibinfo {author} {\bibfnamefont {L.}~\bibnamefont
  {Balicas}}, \bibinfo {author} {\bibfnamefont {R.}~\bibnamefont {Liang}},
  \bibinfo {author} {\bibfnamefont {D.~A.}\ \bibnamefont {Bonn}}, \bibinfo
  {author} {\bibfnamefont {W.~N.}\ \bibnamefont {Hardy}}, \bibinfo {author}
  {\bibfnamefont {C.}~\bibnamefont {Proust}}, \ and\ \bibinfo {author}
  {\bibfnamefont {L.}~\bibnamefont {Taillefer}},\ }\href@noop {} {\bibfield
  {journal} {\bibinfo  {journal} {Phys. Rev. B}\ }\textbf {\bibinfo {volume}
  {83}},\ \bibinfo {pages} {054506} (\bibinfo {year} {2011})}\BibitemShut
  {NoStop}%
\bibitem [{\citenamefont {Doiron-Leyraud}\ \emph {et~al.}(2007)\citenamefont
  {Doiron-Leyraud}, \citenamefont {Proust}, \citenamefont {LeBoeuf},
  \citenamefont {Levallois}, \citenamefont {Bonnemaison}, \citenamefont
  {Liang}, \citenamefont {Bonn}, \citenamefont {Hardy},\ and\ \citenamefont
  {Taillefer}}]{Doiron-Leyraud2007_DPLL}%
  \BibitemOpen
  \bibfield  {author} {\bibinfo {author} {\bibfnamefont {N.}~\bibnamefont
  {Doiron-Leyraud}}, \bibinfo {author} {\bibfnamefont {C.}~\bibnamefont
  {Proust}}, \bibinfo {author} {\bibfnamefont {D.}~\bibnamefont {LeBoeuf}},
  \bibinfo {author} {\bibfnamefont {J.}~\bibnamefont {Levallois}}, \bibinfo
  {author} {\bibfnamefont {J.-B.}\ \bibnamefont {Bonnemaison}}, \bibinfo
  {author} {\bibfnamefont {R.}~\bibnamefont {Liang}}, \bibinfo {author}
  {\bibfnamefont {D.~A.}\ \bibnamefont {Bonn}}, \bibinfo {author}
  {\bibfnamefont {W.~N.}\ \bibnamefont {Hardy}}, \ and\ \bibinfo {author}
  {\bibfnamefont {L.}~\bibnamefont {Taillefer}},\ }\href
  {http://dx.doi.org/10.1038/nature05872} {\bibfield  {journal} {\bibinfo
  {journal} {Nature}\ }\textbf {\bibinfo {volume} {447}},\ \bibinfo {pages}
  {565} (\bibinfo {year} {2007})}\BibitemShut {NoStop}%
\bibitem [{\citenamefont {Monceau}(2012)}]{Monceau2012_Monc}%
  \BibitemOpen
  \bibfield  {author} {\bibinfo {author} {\bibfnamefont {P.}~\bibnamefont
  {Monceau}},\ }\href {\doibase 10.1080/00018732.2012.719674} {\bibfield
  {journal} {\bibinfo  {journal} {Advances in Physics}\ }\textbf {\bibinfo
  {volume} {61}},\ \bibinfo {pages} {325} (\bibinfo {year} {2012})}\BibitemShut
  {NoStop}%
\bibitem [{\citenamefont {Wilson}\ \emph {et~al.}(1975)\citenamefont {Wilson},
  \citenamefont {Di~Salvo},\ and\ \citenamefont {Mahajan}}]{Wilson1975_WiDM}%
  \BibitemOpen
  \bibfield  {author} {\bibinfo {author} {\bibfnamefont {J.}~\bibnamefont
  {Wilson}}, \bibinfo {author} {\bibfnamefont {F.}~\bibnamefont {Di~Salvo}}, \
  and\ \bibinfo {author} {\bibfnamefont {S.}~\bibnamefont {Mahajan}},\ }\href
  {\doibase 10.1080/00018737500101391} {\bibfield  {journal} {\bibinfo
  {journal} {Advances in Physics}\ }\textbf {\bibinfo {volume} {24}},\ \bibinfo
  {pages} {117} (\bibinfo {year} {1975})}\BibitemShut {NoStop}%
\bibitem [{\citenamefont {Woll~Jr.}\ and\ \citenamefont
  {Kohn}(1962)}]{Woll1962_WK}%
  \BibitemOpen
  \bibfield  {author} {\bibinfo {author} {\bibfnamefont {E.~J.}\ \bibnamefont
  {Woll~Jr.}}\ and\ \bibinfo {author} {\bibfnamefont {W.}~\bibnamefont
  {Kohn}},\ }\href@noop {} {\bibfield  {journal} {\bibinfo  {journal} {Phys.
  Rev.}\ }\textbf {\bibinfo {volume} {126}},\ \bibinfo {pages} {1693} (\bibinfo
  {year} {1962})}\BibitemShut {NoStop}%
\bibitem [{\citenamefont {Renker}\ \emph {et~al.}(1973)\citenamefont {Renker},
  \citenamefont {Rietschel}, \citenamefont {Pintschovius}, \citenamefont
  {Glaser}, \citenamefont {Bruesch}, \citenamefont {Kuse},\ and\ \citenamefont
  {Rice}}]{Renker1973_RRPG}%
  \BibitemOpen
  \bibfield  {author} {\bibinfo {author} {\bibfnamefont {B.}~\bibnamefont
  {Renker}}, \bibinfo {author} {\bibfnamefont {H.}~\bibnamefont {Rietschel}},
  \bibinfo {author} {\bibfnamefont {L.}~\bibnamefont {Pintschovius}}, \bibinfo
  {author} {\bibfnamefont {W.}~\bibnamefont {Glaser}}, \bibinfo {author}
  {\bibfnamefont {P.}~\bibnamefont {Bruesch}}, \bibinfo {author} {\bibfnamefont
  {D.}~\bibnamefont {Kuse}}, \ and\ \bibinfo {author} {\bibfnamefont {M.~J.}\
  \bibnamefont {Rice}},\ }\href@noop {} {\bibfield  {journal} {\bibinfo
  {journal} {Phys. Rev. Lett.}\ }\textbf {\bibinfo {volume} {30}},\ \bibinfo
  {pages} {1144} (\bibinfo {year} {1973})}\BibitemShut {NoStop}%
\bibitem [{\citenamefont {Carneiro}\ \emph {et~al.}(1976)\citenamefont
  {Carneiro}, \citenamefont {Shirane}, \citenamefont {Werner},\ and\
  \citenamefont {Kaiser}}]{Carneiro1976_CSWK}%
  \BibitemOpen
  \bibfield  {author} {\bibinfo {author} {\bibfnamefont {K.}~\bibnamefont
  {Carneiro}}, \bibinfo {author} {\bibfnamefont {G.}~\bibnamefont {Shirane}},
  \bibinfo {author} {\bibfnamefont {S.~A.}\ \bibnamefont {Werner}}, \ and\
  \bibinfo {author} {\bibfnamefont {S.}~\bibnamefont {Kaiser}},\ }\href
  {\doibase 10.1103/PhysRevB.13.4258} {\bibfield  {journal} {\bibinfo
  {journal} {Phys. Rev. B}\ }\textbf {\bibinfo {volume} {13}},\ \bibinfo
  {pages} {4258} (\bibinfo {year} {1976})}\BibitemShut {NoStop}%
\bibitem [{\citenamefont {Pouget}\ \emph {et~al.}(1991)\citenamefont {Pouget},
  \citenamefont {Hennion}, \citenamefont {Escribe-Filippini},\ and\
  \citenamefont {Sato}}]{Pouget1991_PHES}%
  \BibitemOpen
  \bibfield  {author} {\bibinfo {author} {\bibfnamefont {J.~P.}\ \bibnamefont
  {Pouget}}, \bibinfo {author} {\bibfnamefont {B.}~\bibnamefont {Hennion}},
  \bibinfo {author} {\bibfnamefont {C.}~\bibnamefont {Escribe-Filippini}}, \
  and\ \bibinfo {author} {\bibfnamefont {M.}~\bibnamefont {Sato}},\ }\href
  {\doibase 10.1103/PhysRevB.43.8421} {\bibfield  {journal} {\bibinfo
  {journal} {Phys. Rev. B}\ }\textbf {\bibinfo {volume} {43}},\ \bibinfo
  {pages} {8421} (\bibinfo {year} {1991})}\BibitemShut {NoStop}%
\bibitem [{\citenamefont {Requardt}\ \emph {et~al.}(2002)\citenamefont
  {Requardt}, \citenamefont {Lorenzo}, \citenamefont {Monceau}, \citenamefont
  {Currat},\ and\ \citenamefont {Krisch}}]{Requardt2002_RLMC}%
  \BibitemOpen
  \bibfield  {author} {\bibinfo {author} {\bibfnamefont {H.}~\bibnamefont
  {Requardt}}, \bibinfo {author} {\bibfnamefont {J.~E.}\ \bibnamefont
  {Lorenzo}}, \bibinfo {author} {\bibfnamefont {P.}~\bibnamefont {Monceau}},
  \bibinfo {author} {\bibfnamefont {R.}~\bibnamefont {Currat}}, \ and\ \bibinfo
  {author} {\bibfnamefont {M.}~\bibnamefont {Krisch}},\ }\href {\doibase
  10.1103/PhysRevB.66.214303} {\bibfield  {journal} {\bibinfo  {journal} {Phys.
  Rev. B}\ }\textbf {\bibinfo {volume} {66}},\ \bibinfo {pages} {214303}
  (\bibinfo {year} {2002})}\BibitemShut {NoStop}%
\bibitem [{\citenamefont {Lorenzo}\ \emph {et~al.}(1998)\citenamefont
  {Lorenzo}, \citenamefont {Currat}, \citenamefont {Monceau}, \citenamefont
  {Hennion}, \citenamefont {Berger},\ and\ \citenamefont
  {Levy}}]{Lorenzo1998_LCMH}%
  \BibitemOpen
  \bibfield  {author} {\bibinfo {author} {\bibfnamefont {J.~E.}\ \bibnamefont
  {Lorenzo}}, \bibinfo {author} {\bibfnamefont {R.}~\bibnamefont {Currat}},
  \bibinfo {author} {\bibfnamefont {P.}~\bibnamefont {Monceau}}, \bibinfo
  {author} {\bibfnamefont {B.}~\bibnamefont {Hennion}}, \bibinfo {author}
  {\bibfnamefont {H.}~\bibnamefont {Berger}}, \ and\ \bibinfo {author}
  {\bibfnamefont {F.}~\bibnamefont {Levy}},\ }\href
  {http://stacks.iop.org/0953-8984/10/i=23/a=010} {\bibfield  {journal}
  {\bibinfo  {journal} {Journal of Physics: Condensed Matter}\ }\textbf
  {\bibinfo {volume} {10}},\ \bibinfo {pages} {5039} (\bibinfo {year}
  {1998})}\BibitemShut {NoStop}%
\bibitem [{\citenamefont {Said}\ \emph {et~al.}(2011)\citenamefont {Said},
  \citenamefont {Sinn},\ and\ \citenamefont {Divan}}]{Said2011_SaSD}%
  \BibitemOpen
  \bibfield  {author} {\bibinfo {author} {\bibfnamefont {A.~H.}\ \bibnamefont
  {Said}}, \bibinfo {author} {\bibfnamefont {H.}~\bibnamefont {Sinn}}, \ and\
  \bibinfo {author} {\bibfnamefont {R.}~\bibnamefont {Divan}},\ }\href
  {\doibase 10.1107/S0909049511001828} {\bibfield  {journal} {\bibinfo
  {journal} {Journal of Synchrotron Radiation}\ }\textbf {\bibinfo {volume}
  {18}},\ \bibinfo {pages} {492} (\bibinfo {year} {2011})}\BibitemShut
  {NoStop}%
\bibitem [{\citenamefont {Toellner}\ \emph {et~al.}(2011)\citenamefont
  {Toellner}, \citenamefont {Alatas},\ and\ \citenamefont
  {Said}}]{Toellner2011_ToAS}%
  \BibitemOpen
  \bibfield  {author} {\bibinfo {author} {\bibfnamefont {T.~S.}\ \bibnamefont
  {Toellner}}, \bibinfo {author} {\bibfnamefont {A.}~\bibnamefont {Alatas}}, \
  and\ \bibinfo {author} {\bibfnamefont {A.~H.}\ \bibnamefont {Said}},\ }\href
  {\doibase 10.1107/S0909049511017535} {\bibfield  {journal} {\bibinfo
  {journal} {Journal of Synchrotron Radiation}\ }\textbf {\bibinfo {volume}
  {18}},\ \bibinfo {pages} {605} (\bibinfo {year} {2011})}\BibitemShut
  {NoStop}%
\bibitem [{\citenamefont {de~Fontaine}\ \emph {et~al.}(1987)\citenamefont
  {de~Fontaine}, \citenamefont {Wille},\ and\ \citenamefont
  {Moss}}]{Fontaine1987_FoWM}%
  \BibitemOpen
  \bibfield  {author} {\bibinfo {author} {\bibfnamefont {D.}~\bibnamefont
  {de~Fontaine}}, \bibinfo {author} {\bibfnamefont {L.~T.}\ \bibnamefont
  {Wille}}, \ and\ \bibinfo {author} {\bibfnamefont {S.~C.}\ \bibnamefont
  {Moss}},\ }\href {\doibase 10.1103/PhysRevB.36.5709} {\bibfield  {journal}
  {\bibinfo  {journal} {Phys. Rev. B}\ }\textbf {\bibinfo {volume} {36}},\
  \bibinfo {pages} {5709} (\bibinfo {year} {1987})}\BibitemShut {NoStop}%
\bibitem [{\citenamefont {Beyers}\ \emph {et~al.}(1989)\citenamefont {Beyers},
  \citenamefont {Ahn}, \citenamefont {Gorman}, \citenamefont {Lee},
  \citenamefont {Parkin}, \citenamefont {Ramirez}, \citenamefont {Roche},
  \citenamefont {Vazquez}, \citenamefont {Gur},\ and\ \citenamefont
  {Huggins}}]{Beyers1989_BAGL}%
  \BibitemOpen
  \bibfield  {author} {\bibinfo {author} {\bibfnamefont {R.}~\bibnamefont
  {Beyers}}, \bibinfo {author} {\bibfnamefont {B.~T.}\ \bibnamefont {Ahn}},
  \bibinfo {author} {\bibfnamefont {G.}~\bibnamefont {Gorman}}, \bibinfo
  {author} {\bibfnamefont {V.~Y.}\ \bibnamefont {Lee}}, \bibinfo {author}
  {\bibfnamefont {S.~S.~P.}\ \bibnamefont {Parkin}}, \bibinfo {author}
  {\bibfnamefont {M.~L.}\ \bibnamefont {Ramirez}}, \bibinfo {author}
  {\bibfnamefont {K.~P.}\ \bibnamefont {Roche}}, \bibinfo {author}
  {\bibfnamefont {J.~E.}\ \bibnamefont {Vazquez}}, \bibinfo {author}
  {\bibfnamefont {T.~M.}\ \bibnamefont {Gur}}, \ and\ \bibinfo {author}
  {\bibfnamefont {R.~A.}\ \bibnamefont {Huggins}},\ }\href
  {http://dx.doi.org/10.1038/340619a0} {\bibfield  {journal} {\bibinfo
  {journal} {Nature}\ }\textbf {\bibinfo {volume} {340}},\ \bibinfo {pages}
  {619} (\bibinfo {year} {1989})}\BibitemShut {NoStop}%
\bibitem [{\citenamefont {v.~Zimmermann}\ \emph {et~al.}(1998)\citenamefont
  {v.~Zimmermann}, \citenamefont {Vigliante}, \citenamefont {Niemoller},
  \citenamefont {Ichikawa}, \citenamefont {Frello}, \citenamefont {Madsen},
  \citenamefont {Wochner}, \citenamefont {Uchida}, \citenamefont {Andersen},
  \citenamefont {Tranquada}, \citenamefont {Gibbs},\ and\ \citenamefont
  {Schneider}}]{Zimmermann1998_ZVNI}%
  \BibitemOpen
  \bibfield  {author} {\bibinfo {author} {\bibfnamefont {M.}~\bibnamefont
  {v.~Zimmermann}}, \bibinfo {author} {\bibfnamefont {A.}~\bibnamefont
  {Vigliante}}, \bibinfo {author} {\bibfnamefont {T.}~\bibnamefont
  {Niemoller}}, \bibinfo {author} {\bibfnamefont {N.}~\bibnamefont {Ichikawa}},
  \bibinfo {author} {\bibfnamefont {T.}~\bibnamefont {Frello}}, \bibinfo
  {author} {\bibfnamefont {J.}~\bibnamefont {Madsen}}, \bibinfo {author}
  {\bibfnamefont {P.}~\bibnamefont {Wochner}}, \bibinfo {author} {\bibfnamefont
  {S.}~\bibnamefont {Uchida}}, \bibinfo {author} {\bibfnamefont {N.~H.}\
  \bibnamefont {Andersen}}, \bibinfo {author} {\bibfnamefont {J.~M.}\
  \bibnamefont {Tranquada}}, \bibinfo {author} {\bibfnamefont {D.}~\bibnamefont
  {Gibbs}}, \ and\ \bibinfo {author} {\bibfnamefont {J.~R.}\ \bibnamefont
  {Schneider}},\ }\href {http://stacks.iop.org/0295-5075/41/i=6/a=629}
  {\bibfield  {journal} {\bibinfo  {journal} {EPL (Europhysics Letters)}\
  }\textbf {\bibinfo {volume} {41}},\ \bibinfo {pages} {629} (\bibinfo {year}
  {1998})}\BibitemShut {NoStop}%
\bibitem [{lif()}]{life_time_note}%
  \BibitemOpen
  \href@noop {} {}\bibinfo {note} {In the SI of
  Ref.~\onlinecite{Chang2012_CBHC}, we argued that the correlation length of
  the CDW $\xi \approx 90$~\AA\ and speed of sound $v_s=4.6 \times
  10^3$~ms$^{-1}$ can also be used to estimate an upper bound energy
  (frequency) scale as $\hbar v_s \xi^{-1} \approx 0.3$~meV.}\BibitemShut
  {Stop}%
\bibitem [{\citenamefont {Timusk}\ and\ \citenamefont
  {Statt}(1999)}]{Timusk1999_TiSt}%
  \BibitemOpen
  \bibfield  {author} {\bibinfo {author} {\bibfnamefont {T.}~\bibnamefont
  {Timusk}}\ and\ \bibinfo {author} {\bibfnamefont {B.}~\bibnamefont {Statt}},\
  }\href@noop {} {\bibfield  {journal} {\bibinfo  {journal} {Rep. Prog. Phys.}\
  }\textbf {\bibinfo {volume} {62}},\ \bibinfo {pages} {61} (\bibinfo {year}
  {1999})}\BibitemShut {NoStop}%
\bibitem [{sam()}]{sample_note}%
  \BibitemOpen
  \href@noop {} {}\bibinfo {note} {Ref.~\onlinecite{Ghiringhelli2012_GLMB}
  studied YBa$_2$Cu$_3$O$_{6.6}$}\BibitemShut {NoStop}%
\bibitem [{\citenamefont {Barmatz}\ \emph {et~al.}(1975)\citenamefont
  {Barmatz}, \citenamefont {Testardi},\ and\ \citenamefont
  {Di~Salvo}}]{Barmatz1975_BTD}%
  \BibitemOpen
  \bibfield  {author} {\bibinfo {author} {\bibfnamefont {M.}~\bibnamefont
  {Barmatz}}, \bibinfo {author} {\bibfnamefont {L.~R.}\ \bibnamefont
  {Testardi}}, \ and\ \bibinfo {author} {\bibfnamefont {F.~J.}\ \bibnamefont
  {Di~Salvo}},\ }\href@noop {} {\bibfield  {journal} {\bibinfo  {journal}
  {Phys. Rev. B}\ }\textbf {\bibinfo {volume} {12}},\ \bibinfo {pages} {4367}
  (\bibinfo {year} {1975})}\BibitemShut {NoStop}%
\bibitem [{\citenamefont {Weber}\ \emph {et~al.}(2011)\citenamefont {Weber},
  \citenamefont {Rosenkranz}, \citenamefont {Castellan}, \citenamefont
  {Osborn}, \citenamefont {Hott}, \citenamefont {Heid}, \citenamefont {Bohnen},
  \citenamefont {Egami}, \citenamefont {Said},\ and\ \citenamefont
  {Reznik}}]{Weber2011_WRCO}%
  \BibitemOpen
  \bibfield  {author} {\bibinfo {author} {\bibfnamefont {F.}~\bibnamefont
  {Weber}}, \bibinfo {author} {\bibfnamefont {S.}~\bibnamefont {Rosenkranz}},
  \bibinfo {author} {\bibfnamefont {J.-P.}\ \bibnamefont {Castellan}}, \bibinfo
  {author} {\bibfnamefont {R.}~\bibnamefont {Osborn}}, \bibinfo {author}
  {\bibfnamefont {R.}~\bibnamefont {Hott}}, \bibinfo {author} {\bibfnamefont
  {R.}~\bibnamefont {Heid}}, \bibinfo {author} {\bibfnamefont {K.-P.}\
  \bibnamefont {Bohnen}}, \bibinfo {author} {\bibfnamefont {T.}~\bibnamefont
  {Egami}}, \bibinfo {author} {\bibfnamefont {A.~H.}\ \bibnamefont {Said}}, \
  and\ \bibinfo {author} {\bibfnamefont {D.}~\bibnamefont {Reznik}},\ }\href
  {\doibase 10.1103/PhysRevLett.107.107403} {\bibfield  {journal} {\bibinfo
  {journal} {Phys. Rev. Lett.}\ }\textbf {\bibinfo {volume} {107}},\ \bibinfo
  {pages} {107403} (\bibinfo {year} {2011})}\BibitemShut {NoStop}%
\bibitem [{\citenamefont {Axe}\ and\ \citenamefont
  {Shirane}(1973)}]{Axe1973_AxSh}%
  \BibitemOpen
  \bibfield  {author} {\bibinfo {author} {\bibfnamefont {J.~D.}\ \bibnamefont
  {Axe}}\ and\ \bibinfo {author} {\bibfnamefont {G.}~\bibnamefont {Shirane}},\
  }\href {\doibase 10.1103/PhysRevB.8.1965} {\bibfield  {journal} {\bibinfo
  {journal} {Phys. Rev. B}\ }\textbf {\bibinfo {volume} {8}},\ \bibinfo {pages}
  {1965} (\bibinfo {year} {1973})}\BibitemShut {NoStop}%
\bibitem [{\citenamefont {Altendorf}\ \emph {et~al.}(1993)\citenamefont
  {Altendorf}, \citenamefont {Chen}, \citenamefont {Irwin}, \citenamefont
  {Liang},\ and\ \citenamefont {Hardy}}]{Altendorf1993_ACIL}%
  \BibitemOpen
  \bibfield  {author} {\bibinfo {author} {\bibfnamefont {E.}~\bibnamefont
  {Altendorf}}, \bibinfo {author} {\bibfnamefont {X.~K.}\ \bibnamefont {Chen}},
  \bibinfo {author} {\bibfnamefont {J.~C.}\ \bibnamefont {Irwin}}, \bibinfo
  {author} {\bibfnamefont {R.}~\bibnamefont {Liang}}, \ and\ \bibinfo {author}
  {\bibfnamefont {W.~N.}\ \bibnamefont {Hardy}},\ }\href {\doibase
  10.1103/PhysRevB.47.8140} {\bibfield  {journal} {\bibinfo  {journal} {Phys.
  Rev. B}\ }\textbf {\bibinfo {volume} {47}},\ \bibinfo {pages} {8140}
  (\bibinfo {year} {1993})}\BibitemShut {NoStop}%
\bibitem [{\citenamefont {Reznik}(2012)}]{Reznik2012_Rezn}%
  \BibitemOpen
  \bibfield  {author} {\bibinfo {author} {\bibfnamefont {D.}~\bibnamefont
  {Reznik}},\ }\bibfield  {booktitle} {\emph {\bibinfo {booktitle} {Stripes and
  Electronic Liquid Crystals in Strongly Correlated Materials}},\ }\href
  {http://www.sciencedirect.com/science/article/pii/S0921453412000469}
  {\bibfield  {journal} {\bibinfo  {journal} {Physica C: Superconductivity}\
  }\textbf {\bibinfo {volume} {481}},\ \bibinfo {pages} {75} (\bibinfo {year}
  {2012})}\BibitemShut {NoStop}%
\bibitem [{\citenamefont {Lee}\ \emph {et~al.}(2007)\citenamefont {Lee},
  \citenamefont {Vishik}, \citenamefont {Tanaka}, \citenamefont {Lu},
  \citenamefont {Sasagawa}, \citenamefont {Nagaosa}, \citenamefont {Devereaux},
  \citenamefont {Hussain},\ and\ \citenamefont {Shen}}]{Lee2007_LVTL}%
  \BibitemOpen
  \bibfield  {author} {\bibinfo {author} {\bibfnamefont {W.~S.}\ \bibnamefont
  {Lee}}, \bibinfo {author} {\bibfnamefont {I.~M.}\ \bibnamefont {Vishik}},
  \bibinfo {author} {\bibfnamefont {K.}~\bibnamefont {Tanaka}}, \bibinfo
  {author} {\bibfnamefont {D.~H.}\ \bibnamefont {Lu}}, \bibinfo {author}
  {\bibfnamefont {T.}~\bibnamefont {Sasagawa}}, \bibinfo {author}
  {\bibfnamefont {N.}~\bibnamefont {Nagaosa}}, \bibinfo {author} {\bibfnamefont
  {T.~P.}\ \bibnamefont {Devereaux}}, \bibinfo {author} {\bibfnamefont
  {Z.}~\bibnamefont {Hussain}}, \ and\ \bibinfo {author} {\bibfnamefont
  {Z.-X.}\ \bibnamefont {Shen}},\ }\href
  {http://dx.doi.org/10.1038/nature06219} {\bibfield  {journal} {\bibinfo
  {journal} {Nature}\ }\textbf {\bibinfo {volume} {450}},\ \bibinfo {pages}
  {81} (\bibinfo {year} {2007})}\BibitemShut {NoStop}%
\bibitem [{\citenamefont {Hinton}\ \emph {et~al.}()\citenamefont {Hinton},
  \citenamefont {Koralek}, \citenamefont {Lu}, \citenamefont {Vishwanath},
  \citenamefont {Orenstein}, \citenamefont {Bonn}, \citenamefont {Hardy},\ and\
  \citenamefont {Liang}}]{Hinton_HKLV}%
  \BibitemOpen
  \bibfield  {author} {\bibinfo {author} {\bibfnamefont {J.~P.}\ \bibnamefont
  {Hinton}}, \bibinfo {author} {\bibfnamefont {J.~D.}\ \bibnamefont {Koralek}},
  \bibinfo {author} {\bibfnamefont {Y.~M.}\ \bibnamefont {Lu}}, \bibinfo
  {author} {\bibfnamefont {A.}~\bibnamefont {Vishwanath}}, \bibinfo {author}
  {\bibfnamefont {J.}~\bibnamefont {Orenstein}}, \bibinfo {author}
  {\bibfnamefont {D.~A.}\ \bibnamefont {Bonn}}, \bibinfo {author}
  {\bibfnamefont {W.~N.}\ \bibnamefont {Hardy}}, \ and\ \bibinfo {author}
  {\bibfnamefont {R.}~\bibnamefont {Liang}},\ }\href@noop {} {}\Eprint
  {http://arxiv.org/abs/arXiv:1305.1361} {arXiv:1305.1361} \BibitemShut
  {NoStop}%
\bibitem [{\citenamefont {Tacon}\ \emph {et~al.}()\citenamefont {Tacon},
  \citenamefont {Bosak}, \citenamefont {Souliou}, \citenamefont {Dellea},
  \citenamefont {Loew}, \citenamefont {Heid}, \citenamefont {Bohnen},
  \citenamefont {Ghiringhelli}, \citenamefont {Krisch},\ and\ \citenamefont
  {Keimer}}]{Tacon2013_TBSD}%
  \BibitemOpen
  \bibfield  {author} {\bibinfo {author} {\bibfnamefont {M.~L.}\ \bibnamefont
  {Tacon}}, \bibinfo {author} {\bibfnamefont {A.}~\bibnamefont {Bosak}},
  \bibinfo {author} {\bibfnamefont {S.~M.}\ \bibnamefont {Souliou}}, \bibinfo
  {author} {\bibfnamefont {G.}~\bibnamefont {Dellea}}, \bibinfo {author}
  {\bibfnamefont {T.}~\bibnamefont {Loew}}, \bibinfo {author} {\bibfnamefont
  {R.}~\bibnamefont {Heid}}, \bibinfo {author} {\bibfnamefont {K.-P.}\
  \bibnamefont {Bohnen}}, \bibinfo {author} {\bibfnamefont {G.}~\bibnamefont
  {Ghiringhelli}}, \bibinfo {author} {\bibfnamefont {M.}~\bibnamefont
  {Krisch}}, \ and\ \bibinfo {author} {\bibfnamefont {B.}~\bibnamefont
  {Keimer}},\ }\href@noop {} {}\Eprint {http://arxiv.org/abs/arXiv:1307.1673}
  {arXiv:1307.1673} \BibitemShut {NoStop}%
\bibitem [{\citenamefont {Raichle}\ \emph {et~al.}(2011)\citenamefont
  {Raichle}, \citenamefont {Reznik}, \citenamefont {Lamago}, \citenamefont
  {Heid}, \citenamefont {Li}, \citenamefont {Bakr}, \citenamefont {Ulrich},
  \citenamefont {Hinkov}, \citenamefont {Hradil}, \citenamefont {Lin},\ and\
  \citenamefont {Keimer}}]{Raichle2011_RRLH}%
  \BibitemOpen
  \bibfield  {author} {\bibinfo {author} {\bibfnamefont {M.}~\bibnamefont
  {Raichle}}, \bibinfo {author} {\bibfnamefont {D.}~\bibnamefont {Reznik}},
  \bibinfo {author} {\bibfnamefont {D.}~\bibnamefont {Lamago}}, \bibinfo
  {author} {\bibfnamefont {R.}~\bibnamefont {Heid}}, \bibinfo {author}
  {\bibfnamefont {Y.}~\bibnamefont {Li}}, \bibinfo {author} {\bibfnamefont
  {M.}~\bibnamefont {Bakr}}, \bibinfo {author} {\bibfnamefont {C.}~\bibnamefont
  {Ulrich}}, \bibinfo {author} {\bibfnamefont {V.}~\bibnamefont {Hinkov}},
  \bibinfo {author} {\bibfnamefont {K.}~\bibnamefont {Hradil}}, \bibinfo
  {author} {\bibfnamefont {C.~T.}\ \bibnamefont {Lin}}, \ and\ \bibinfo
  {author} {\bibfnamefont {B.}~\bibnamefont {Keimer}},\ }\href {\doibase
  10.1103/PhysRevLett.107.177004} {\bibfield  {journal} {\bibinfo  {journal}
  {Phys. Rev. Lett.}\ }\textbf {\bibinfo {volume} {107}},\ \bibinfo {pages}
  {177004} (\bibinfo {year} {2011})}\BibitemShut {NoStop}%
\bibitem [{\citenamefont {Reznik}\ \emph {et~al.}(2008)\citenamefont {Reznik},
  \citenamefont {Pintschovius}, \citenamefont {Tranquada}, \citenamefont
  {Arai}, \citenamefont {Endoh}, \citenamefont {Masui},\ and\ \citenamefont
  {Tajima}}]{Reznik2008_RPTA}%
  \BibitemOpen
  \bibfield  {author} {\bibinfo {author} {\bibfnamefont {D.}~\bibnamefont
  {Reznik}}, \bibinfo {author} {\bibfnamefont {L.}~\bibnamefont
  {Pintschovius}}, \bibinfo {author} {\bibfnamefont {J.~M.}\ \bibnamefont
  {Tranquada}}, \bibinfo {author} {\bibfnamefont {M.}~\bibnamefont {Arai}},
  \bibinfo {author} {\bibfnamefont {Y.}~\bibnamefont {Endoh}}, \bibinfo
  {author} {\bibfnamefont {T.}~\bibnamefont {Masui}}, \ and\ \bibinfo {author}
  {\bibfnamefont {S.}~\bibnamefont {Tajima}},\ }\href {\doibase
  10.1103/PhysRevB.78.094507} {\bibfield  {journal} {\bibinfo  {journal} {Phys.
  Rev. B}\ }\textbf {\bibinfo {volume} {78}},\ \bibinfo {pages} {094507}
  (\bibinfo {year} {2008})}\BibitemShut {NoStop}%
\bibitem [{\citenamefont {Chung}\ \emph {et~al.}(2003)\citenamefont {Chung},
  \citenamefont {Egami}, \citenamefont {McQueeney}, \citenamefont {Yethiraj},
  \citenamefont {Arai}, \citenamefont {Yokoo}, \citenamefont {Petrov},
  \citenamefont {Mook}, \citenamefont {Endoh}, \citenamefont {Tajima},
  \citenamefont {Frost},\ and\ \citenamefont {Dogan}}]{Chung2003_CEMY}%
  \BibitemOpen
  \bibfield  {author} {\bibinfo {author} {\bibfnamefont {J.-H.}\ \bibnamefont
  {Chung}}, \bibinfo {author} {\bibfnamefont {T.}~\bibnamefont {Egami}},
  \bibinfo {author} {\bibfnamefont {R.~J.}\ \bibnamefont {McQueeney}}, \bibinfo
  {author} {\bibfnamefont {M.}~\bibnamefont {Yethiraj}}, \bibinfo {author}
  {\bibfnamefont {M.}~\bibnamefont {Arai}}, \bibinfo {author} {\bibfnamefont
  {T.}~\bibnamefont {Yokoo}}, \bibinfo {author} {\bibfnamefont
  {Y.}~\bibnamefont {Petrov}}, \bibinfo {author} {\bibfnamefont {H.~A.}\
  \bibnamefont {Mook}}, \bibinfo {author} {\bibfnamefont {Y.}~\bibnamefont
  {Endoh}}, \bibinfo {author} {\bibfnamefont {S.}~\bibnamefont {Tajima}},
  \bibinfo {author} {\bibfnamefont {C.}~\bibnamefont {Frost}}, \ and\ \bibinfo
  {author} {\bibfnamefont {F.}~\bibnamefont {Dogan}},\ }\href {\doibase
  10.1103/PhysRevB.67.014517} {\bibfield  {journal} {\bibinfo  {journal} {Phys.
  Rev. B}\ }\textbf {\bibinfo {volume} {67}},\ \bibinfo {pages} {014517}
  (\bibinfo {year} {2003})}\BibitemShut {NoStop}%
\bibitem [{\citenamefont {Stercel}\ \emph {et~al.}(2008)\citenamefont
  {Stercel}, \citenamefont {Egami}, \citenamefont {Mook}, \citenamefont
  {Yethiraj}, \citenamefont {Chung}, \citenamefont {Arai}, \citenamefont
  {Frost},\ and\ \citenamefont {Dogan}}]{Stercel2008_SEMY}%
  \BibitemOpen
  \bibfield  {author} {\bibinfo {author} {\bibfnamefont {F.}~\bibnamefont
  {Stercel}}, \bibinfo {author} {\bibfnamefont {T.}~\bibnamefont {Egami}},
  \bibinfo {author} {\bibfnamefont {H.~A.}\ \bibnamefont {Mook}}, \bibinfo
  {author} {\bibfnamefont {M.}~\bibnamefont {Yethiraj}}, \bibinfo {author}
  {\bibfnamefont {J.-H.}\ \bibnamefont {Chung}}, \bibinfo {author}
  {\bibfnamefont {M.}~\bibnamefont {Arai}}, \bibinfo {author} {\bibfnamefont
  {C.}~\bibnamefont {Frost}}, \ and\ \bibinfo {author} {\bibfnamefont
  {F.}~\bibnamefont {Dogan}},\ }\href {\doibase 10.1103/PhysRevB.77.014502}
  {\bibfield  {journal} {\bibinfo  {journal} {Phys. Rev. B}\ }\textbf {\bibinfo
  {volume} {77}},\ \bibinfo {pages} {014502} (\bibinfo {year}
  {2008})}\BibitemShut {NoStop}%
\bibitem [{\citenamefont {Pintschovius}\ and\ \citenamefont
  {Braden}(1999)}]{Pintschovius1999_PiBr}%
  \BibitemOpen
  \bibfield  {author} {\bibinfo {author} {\bibfnamefont {L.}~\bibnamefont
  {Pintschovius}}\ and\ \bibinfo {author} {\bibfnamefont {M.}~\bibnamefont
  {Braden}},\ }\href {\doibase 10.1103/PhysRevB.60.R15039} {\bibfield
  {journal} {\bibinfo  {journal} {Phys. Rev. B}\ }\textbf {\bibinfo {volume}
  {60}},\ \bibinfo {pages} {R15039} (\bibinfo {year} {1999})}\BibitemShut
  {NoStop}%
\bibitem [{\citenamefont {Schmidt}\ \emph {et~al.}(2011)\citenamefont
  {Schmidt}, \citenamefont {Fujita}, \citenamefont {Kim}, \citenamefont
  {Lawler}, \citenamefont {Eisaki}, \citenamefont {Uchida}, \citenamefont
  {Lee},\ and\ \citenamefont {Davis}}]{Schmidt2011_SFKL}%
  \BibitemOpen
  \bibfield  {author} {\bibinfo {author} {\bibfnamefont {A.~R.}\ \bibnamefont
  {Schmidt}}, \bibinfo {author} {\bibfnamefont {K.}~\bibnamefont {Fujita}},
  \bibinfo {author} {\bibfnamefont {E.-A.}\ \bibnamefont {Kim}}, \bibinfo
  {author} {\bibfnamefont {M.~J.}\ \bibnamefont {Lawler}}, \bibinfo {author}
  {\bibfnamefont {H.}~\bibnamefont {Eisaki}}, \bibinfo {author} {\bibfnamefont
  {S.}~\bibnamefont {Uchida}}, \bibinfo {author} {\bibfnamefont {D.-H.}\
  \bibnamefont {Lee}}, \ and\ \bibinfo {author} {\bibfnamefont {J.~C.}\
  \bibnamefont {Davis}},\ }\href
  {http://stacks.iop.org/1367-2630/13/i=6/a=065014} {\bibfield  {journal}
  {\bibinfo  {journal} {New Journal of Physics}\ }\textbf {\bibinfo {volume}
  {13}},\ \bibinfo {pages} {065014} (\bibinfo {year} {2011})}\BibitemShut
  {NoStop}%
\bibitem [{\citenamefont {da~Silva~Neto}\ \emph {et~al.}(2012)\citenamefont
  {da~Silva~Neto}, \citenamefont {Aynajian}, \citenamefont {Parker},\ and\
  \citenamefont {Yazdani}}]{SilvaNeto2012_SAPY}%
  \BibitemOpen
  \bibfield  {author} {\bibinfo {author} {\bibfnamefont {E.~H.}\ \bibnamefont
  {da~Silva~Neto}}, \bibinfo {author} {\bibfnamefont {P.}~\bibnamefont
  {Aynajian}}, \bibinfo {author} {\bibfnamefont {C.~V.}\ \bibnamefont
  {Parker}}, \ and\ \bibinfo {author} {\bibfnamefont {A.}~\bibnamefont
  {Yazdani}},\ }\bibfield  {booktitle} {\emph {\bibinfo {booktitle} {Stripes
  and Electronic Liquid Crystals in Strongly Correlated Materials}},\ }\href
  {http://www.sciencedirect.com/science/article/pii/S0921453412002055}
  {\bibfield  {journal} {\bibinfo  {journal} {Physica C: Superconductivity}\
  }\textbf {\bibinfo {volume} {481}},\ \bibinfo {pages} {153} (\bibinfo {year}
  {2012})}\BibitemShut {NoStop}%
\bibitem [{\citenamefont {Chan}\ and\ \citenamefont
  {Heine}(1973)}]{Chan1973_ChHe}%
  \BibitemOpen
  \bibfield  {author} {\bibinfo {author} {\bibfnamefont {S.~K.}\ \bibnamefont
  {Chan}}\ and\ \bibinfo {author} {\bibfnamefont {V.}~\bibnamefont {Heine}},\
  }\href {http://stacks.iop.org/0305-4608/3/i=4/a=022} {\bibfield  {journal}
  {\bibinfo  {journal} {J. Phys. F}\ }\textbf {\bibinfo {volume} {3}},\
  \bibinfo {pages} {795} (\bibinfo {year} {1973})}\BibitemShut {NoStop}%
\end{thebibliography}%
\bibliographystyle{apsrev4-1}

\end{document}